\documentclass[12pt]{article}
\usepackage{amsfonts}
\usepackage{amssymb}
\usepackage[dvips]{graphics}
\newcommand{\Nl}{\mathbb{N}}
\newcommand{\Ir}{\mathbb{Z}}
\newcommand{\Cx}{\mathbb{C}}

\newtheorem{theorem}{Theorem}[section]
          
\newtheorem{lemma}[theorem]{Lemma}              
\newtheorem{proposition}[theorem]{Proposition}
\newtheorem{corollary}[theorem]{Corollary}
\newtheorem{definition}[theorem]{Definition}

\newcommand{\rem}[1]{{\bf Remark:}}
\newcommand{\condmat}[1]{archived as {\tt cond-mat/#1}}

\renewcommand{\theequation}{\thesection.\arabic{equation}}
\newcommand{\eq}[1]{(\ref{#1})}
\newenvironment{proof}{\noindent {\bf Proof: }}{\QED\medskip}
\def\QED{{\hspace*{\fill}{\vrule height .5ex width 1ex }\quad} 
    \vskip 0pt plus20pt}
\newcommand{\be}{\begin{equation}}
\newcommand{\ee}{\end{equation}}
\newcommand{\bea}{\begin{eqnarray}}
\newcommand{\eea}{\end{eqnarray}}
\newcommand{\beann}{\begin{eqnarray*}}
\newcommand{\eeann}{\end{eqnarray*}}

\newcommand{\ket}[1]{\vert{#1}\rangle}
\newcommand{\bra}[1]{\langle{#1}\vert}

\newcommand{\unity}{{1\hskip -3pt \rm{I}}}
\newcommand{\ip}[2]{\langle{#1|#2}\rangle}
\newcommand{\Rayleigh}[2]{\frac{\ip{#1}{#2 #1}}{\ip{#1}{#1}}}
\newcommand{\Hil}{\mathcal{H}}
\newcommand{\calK}{\mathcal{K}}
\newcommand{\Hpp}{H^{++}}
\newcommand{\HXXZ}{H^{\rm{XXZ}}}

\newcommand{\allup}{\ket{\uparrow\dots\uparrow}}
\newcommand{\alldown}{\ket{\downarrow\dots\downarrow}}
\newcommand{\Proj}{{\rm Proj}}
\newcommand{\Span}{{\rm span}}
\newcommand{\binom}[2]{\left(\hspace{-3pt}
	\begin{array}{c}#1 \\ #2\end{array}\hspace{-3pt}\right)}
\newcommand{\sbinom}[2]{\scriptstyle{\left(\hspace{-3pt}
	\begin{array}{c}\scriptstyle #1 \\ 
        \scriptstyle #2\end{array}\hspace{-3pt}\right)}}
\newcommand{\qbinom}[3]{{\left[\hspace{-3pt}
	\begin{array}{c}#1 \\ #2\end{array}\hspace{-3pt}\right]_{#3}}}
\newcommand{\sqbinom}[3]{\left[\hspace{-3pt}
	\begin{array}{c}\scriptstyle #1 \\ 
        \scriptstyle #2\end{array}\hspace{-3pt}\right]_{#3}}
\newcommand{\pn}[1]{f_q({#1})}

\newcommand{\floor}[1]{\left\lfloor{#1}\right\rfloor}
\newcommand{\ceil}[1]{\left\lceil{#1}\right\rceil}
\newcommand{\EXP}[2]{{\langle{#1}\rangle}_{#2}}

\begin{document}
{\baselineskip=10pt \thispagestyle{empty} {{\small Preprint UC Davis Math
2000-16}
\hspace{\fill}}

\vspace{20pt}

\begin{center}
{\LARGE \bf Droplet States in the XXZ Heisenberg Chain\\[27pt]}
{\large \bf Bruno Nachtergaele 
and Shannon Starr\\[10pt]}
{\large  Department of Mathematics\\
University of California, Davis\\
Davis, CA 95616-8633, USA\\[15pt]}
{\normalsize bxn@math.ucdavis.edu, sstarr@math.ucdavis.edu}\\[30pt]
%(date)\\[30pt]
\end{center}

{\bf Abstract:}
We consider the ground states of the ferromagnetic XXZ chain with 
spin up boundary conditions in sectors with a fixed number of down
spins. This forces the existence of a droplet of down spins in the
system. We find the exact energy and the states that describe these
droplets in the limit of an infinite number of down spins. We prove 
that there is a gap in the spectrum above the droplet states. 
As the XXZ Hamiltonian has a gap above the fully magnetized ground 
states as well, this means that the droplet states (for sufficiently
large droplets) form an isolated band. The width of this band tends
to zero in the limit of infinitely large droplets.
We also prove the analogous results for finite chains with periodic 
boundary conditions and for the infinite chain.

\vspace{8pt}
{\small \bf Keywords:} Anisotropic Heisenberg ferromagnet, XXZ chain,
droplet states, excitations, spectral gap.
\vskip .2 cm
\noindent
{\small \bf PACS 1999 numbers:} 05.70.Np, 75.10.Jm, 75.30.Kz, 75.70.Kw 
\newline
{\small \bf MCS 2000 numbers:} 82B10, 82B24, 82D40 
\vfill
\hrule width2truein \smallskip {\baselineskip=10pt \noindent Copyright
\copyright\ 2000 by the authors. Reproduction of this article in its entirety, 
by any means, is permitted for non-commercial purposes.\par }}

%%%%%%%%%%%%%%%%%%%%%%%%%%%%%%%%%%%%%%%%%%%%%%%%%%%%%%%%%%%%%%%%%%%
%%%%%%%%%%%%%%%%%%%%%%%%%%%%%%%%%%%%%%%%%%%%%%%%%%%%%%%%%%%%%%%%%%%
%%%%%%%%%%%%%%%%%%%%%%%%%%%%%%%%%%%%%%%%%%%%%%%%%%%%%%%%%%%%%%%%%%%
\newpage
\section{Introduction}

Droplet states have been studied in considerable detail for the Ising model
\cite{DKS,Pfi,BIV}, where they play an important role in understanding 
dynamical phenomena \cite{SS}. In this paper we  consider the
spin-$\frac{1}{2}$ ferromagnetic XXZ Heisenberg chain and prove that the bottom
of its spectrum consists of an isolated nearly flat band of droplet states in 
a sense made precise below.

The Hamiltonian for a chain of $L$ spins acts on the Hilbert space
$$
  \Hil_L = \Cx_1^2 \otimes \dots \otimes \Cx_L^2
$$
as the sum of nearest-neighbor interactions
$$
  \HXXZ_{[1,L]} = \sum_{x=1}^{L-1} \HXXZ_{x,x+1}\, 
$$
of the form
\begin{equation}
\label{XXZ:Ham}
\HXXZ_{x,x+1} 
  = - \Delta^{-1} (\vec{S}_x \cdot \vec{S}_{x+1} - \frac{1}{4})
  - (1 - \Delta^{-1}) (S_x^3 S_{x+1}^3 - \frac{1}{4})\, .
\end{equation}
Here $S_x^i$ ($i=1,2,3$) are the spin matrices,
acting on $\Cx^2_x$, extended by unity to $\Hil_L$,
and normalized so that they have eigenvalues $\pm 1/2$.
The anisotropy parameter, $\Delta$, is always assumed to be
$> 1$. To formulate the results and also for the proofs,
we need to consider the following combinations of boundary fields for
systems defined on an arbitrary interval:
for $\alpha,\beta=\pm 1,0$, and $[a,b]\subset\Ir$, define
\begin{equation}
H^{\alpha\beta}_{[a,b]}= \sum_{x=a}^{b-1}\HXXZ_{x,x+1} - A(\Delta)(\alpha
S^{3}_a+\beta S^{3}_b)\quad, 
\label{Hab}\end{equation}
where $A(\Delta) = \frac{1}{2} \sqrt{1 - \Delta^{-2}}$. Note that
$H^{00}_{[1,L]}=\HXXZ_{[1,L]}$.

As all the Hamiltonians $H^{\alpha\beta}_{[a,b]}$ commute with the 
total third component of the spin, it makes sense to study their
ground states restricted to a subspace of fixed number of down spins. 
The subspace for a chain of $L$ spins consisting of the states with
$n$ down spins will be denoted by $\Hil_{L,n}$, for $0\leq n\leq L$.
In all cases the ground state is then unique. 
The Hamiltonians with $+-$ and $-+$ boundary fields have been studied 
extensively and have kink and antikink ground states respectively 
\cite{ASW,GW,KN1,Mat,KN3,BCN,BM}. The unique ground states for a chain
on $[a,b]\subset\Ir$, in the sector with $n$ down spins, will be
denoted by $\psi^{\alpha\beta}_{[a,b]}(n), 0\leq n\leq b-a+1$.
For $\alpha\beta=+-, -+$, they are given by 
\begin{eqnarray}
\label{Intro:+-}
\psi^{+-}_{[a,b]}(n) = 
  \sum_{a \leq x_1 < \dots < x_n \leq b} q^{\sum_{k=1}^n (b+1-x_k)}
  \left( \prod_{k=1}^n S_{x_k}^- \right) \ket{\uparrow \dots 
  \uparrow}_{[a,b]} \\
\label{Intro:-+}
\psi^{-+}_{[a,b]}(n) = 
  \sum_{a \leq x_1 < \dots < x_n \leq b} q^{\sum_{k=1}^n (x_k+1-a)}
  \left( \prod_{k=1}^n S_{x_k}^- \right) \ket{\uparrow \dots \uparrow}_{[a,b]} 
\end{eqnarray}
where $\Delta = (q+q^{-1})/2$.
Note that the norm of these vectors depends on the length (but not on the
position) of the interval $[a,b]$ (see \eq{App:kink-norm}). There is a uniform 
lower bound for the spectral gap above these ground states \cite{KN1}, a 
property that will be essential in the proofs. 

Here, we are interested in the ground states of the Hamiltonian
with $++$ boundary fields, which we refer to as the {\it droplet
Hamiltonian}, in the regime where there are a sufficently large number
of down spins. This includes, but is not limited 
to, the case where there is a fixed density $\rho$, $0<\rho\leq 1$,
of down spins in a system with $++$ boundary conditions. We 
prove that under these conditions the ground states contain one droplet of
down spins in a background of up spins.

From the mathematical point-of-view there is an important
distinction between the kink Hamiltonian and the droplet Hamiltonian,
which is that the droplet Hamiltonian does not possess $SU_q(2)$ symmetry. 
In contrast to the kink Hamiltonian where explicit formulae
are known for the ground states in finite volumes, no such explicit
analytic formulae are known for the droplet Hamiltonian for general $L$.
Therefore, we rely primarily on energy estimates, and our main
results are formulated as estimates that become exact only in 
the limit $n, L\to\infty$. This is natural as, again unlike for the kink 
ground states, there is no immediate infinite-volume description of the 
droplet states. We find the exact energy of an infinite droplet and an 
approximation of the droplet ground states that becomes exact in the 
thermodynamic limit. We also prove that all states with the energy of the
droplet are necessarily droplet states, again, in the thermodynamic limit. For
the droplet Hamiltonians this means that the droplet states are all the ground
states, and that there is a gap above them. One can also interpret this as
saying that all excitations of the fully magnetized ground states of the XXZ
chain, with sufficiently many overturned spins and not too high an energy, are
droplet states.

\subsection{Main Result}
The main result of this paper is the approximate
calculation of the ground state
energy, the ground state space, and a lower bound for the
spectral gap of the operator
$H^{++}_{[1,L]}$ restricted to the sector $\Hil_{L,n}$.
If the results were exact, we would have an eigenvalue $E_0$, a subspace
$\Hil_{L,n}^0 \subset \Hil_{L,n}$, and a positive number $\gamma$, such that
$$
H^{++}_{[1,L]} \textrm{Proj}(\Hil^0_{L,n}) = E_0 \textrm{Proj}(\Hil^0_{L,n})
$$
and
$$
H^{++}_{[1,L]} \textrm{Proj}(\Hil_{L,n})
  \geq E_0 \textrm{Proj}(\Hil_{L,n}) + \gamma (\textrm{Proj}(\Hil_{L,n})
  - \textrm{Proj}(\Hil^0_{L,n}))\, .
$$
We will always use the notation ${\rm Proj}(V)$ to mean orthogonal projection
onto a subspace $V$.

\begin{figure}
\begin{center}
\resizebox{12truecm}{2truecm}{\includegraphics{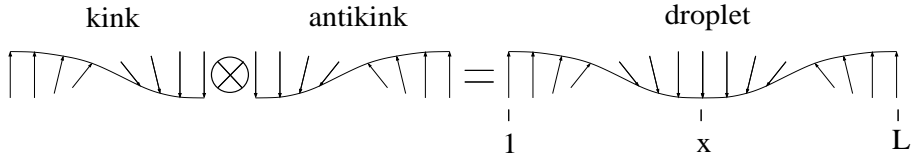}}
\parbox{11truecm}{\caption{\baselineskip=5 pt\small
\label{Fig:typical-drop}
Diagram of a typical droplet as the tensor product of a kink
and antikink.}
}
\end{center}
\end{figure}
Our results are approximations, with increasing accuracy as
$n$ tends to infinity, independent of $L$.
First, we identify the proposed ground state space.
For $n\geq 0$ and $\floor{n/2} \leq x \leq L - \ceil{n/2}$
define
\be
\xi_{L,n}(x) \
  = \psi^{+-}_{[1,x]}(\floor{n/2})
  \otimes \psi^{-+}_{[x+1,L]}(\ceil{n/2})\, .
\label{def_drop}\ee
For any real number $x$, $\floor{x}$ is the greatest integer $\leq x$,
and $\ceil{x}$ is the least integer $\geq x$.
The typical magnetization profile of $\xi_{L,n}(x)$ is shown in 
Figure \ref{Fig:typical-drop}.
We define the space of approximate ground states as follows: 
$$
\calK_{L,n} = \Span \{ \xi_{L,n}(x) : \, 
\floor{n/2}\leq x \leq L-\ceil{n/2} \}\, .
$$
$\calK_{L,n}$ is the space of ``approximate'' droplet states 
with $n$ down spins for a finite chain of length $L$.
An interval of length $n$ can occur in $L-n+1$ positions inside 
a chain of length $L$.
This explains why $\dim \calK_{L,n} = L-n+1$.

Alternatively, we could use the following definitions of
approximate droplet states:
$$
\xi'_{L,n}(x)=[S^{\rm{antikink},+}_{[1,L]}]^{x-\floor{n/2}}
[S^{\rm{kink},+}_{[1,L]}]^{L-\ceil{n/2}-x}\alldown
$$
where $S^{\rm{kink},+}_{[1,L]}$ is the $SU_q(2)$ raising operator
(see, e.g., (2.5b) of \cite{KN1}), and $S^{\rm{antikink},+}_{[1,L]}$ 
is the left-right reflection of $S^{\rm{kink},+}_{[1,L]}$. Yet another
option for the droplet states is to take the exact ground states of
the Hamiltonians $H_{[1,L]}=H_{[1,x]}^{+-}+ H_{[x,L]}^{-+}$, which
have a pinning field at position $x$, and for which exact expressions
for the ground states can be obtained. One can show that suitable linear
combinations of these states differ in norm from the $\xi_{L,n}(x)$ by 
no more
than $O(q^n)$. We will only use the states $\xi_{L,n}(x)$ defined in 
\eq{def_drop}, as they have a more intuitive interpretation as a tensor
product of a kink and an antikink state.

\begin{theorem} $\vspace{1mm}$
\label{main:theorem}

a) There exists a constant $C < \infty$ such that
$$
\| (H^{++}_{[1,L]} - A(\Delta)) \Proj(\calK_{L,n}) \|
  \leq C q^n\, .
$$
The constant $C$ depends only on $q$, not on $L$ or $n$.

b) There exists a sequence $\epsilon_n$,
with $\lim_{n \to \infty} \epsilon_n = 0$,
such that
\begin{eqnarray*}
&&
H^{++}_{[1,L]} \Proj(\Hil_{L,n}) \geq (A(\Delta) - 2 C q^n) \Proj(\Hil_{L,n}) \\
&& \hspace{4cm}
  + (\gamma - \epsilon_n)(\Proj(\Hil_{L,n}) - \Proj(\calK_{L,n}))\, ,
\end{eqnarray*}
where $\gamma = 1 - \Delta^{-1}$.
The sequence $\epsilon_n$ can be chosen to decay at least as fast as $n^{-1/4}$,
independent of $L$.
\end{theorem}

For $H^{XXZ}_{[1,L]}$, which is the one without boundary terms, the
large-droplet states are not separated in the spectrum from other excitations
such as the spin waves, i.e., the band of continuous spectrum due to spin wave
excitations overlaps with the states of droplet type.
Although similar results should hold for boundary fields of larger magnitude
the value, $A(\Delta)$, of the boundary fields in the droplet
Hamiltonian, is particularly convenient for at least two reasons: 1)
it allows us to write the Hamiltonian as a sum of kink and anti-kink
Hamiltonians, which is the basis for many of our arguments, 2) the
energy of a droplet in the center of the chain is the same as for a
droplet attached to the boundary. This allows us to construct explicitly
the subspace of all droplet states asymptotically in the thermodynamic
limit.

\begin{figure}
\begin{center}
\resizebox{12truecm}{6truecm}{\includegraphics{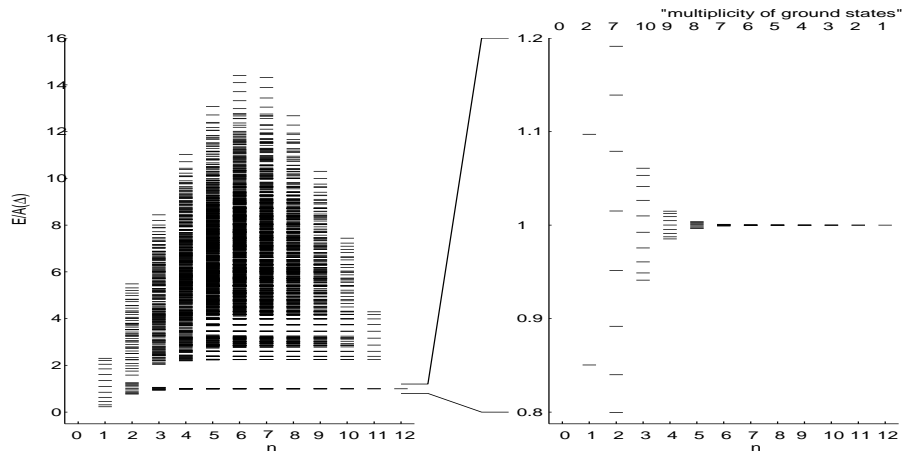}}
\parbox{11truecm}{\caption{\baselineskip=5 pt\small
\label{fig:dropspec}
Spectrum for $H^{++}_{[1,12]}$ when $\Delta = 2.125$ ($q=1/4$).}
}
\end{center}
\end{figure}

Although our main results are about infinite droplets, i.e., they are 
asymptotic properties of finite droplets in the limit of their size
tending to infinity, we can extract from our proofs estimates of the 
corrections for finite size droplets. This allows the following reformulation
of the main result in terms of the eigenvalues near the bottom
of the spectrum and the corresponding eigenprojection.
Let 
$\lambda_{L,n}(1) \leq \lambda_{L,n}(2) \leq \dots $
be the eigenvalues of $H^{++}_{[1,L]}$ restricted to the sector $\Hil_{L,n}$.
Let $\psi^{++}_{L,n}(1), \psi^{++}_{L,n}(2), \dots $ 
be the corresponding eigenstates, and define
$$
  \Hil^{k}_{L,n} = \Span \{ \psi^{++}_{L,n}(j) : 1 \leq j \leq k\}\, .
$$
\begin{theorem} 
\label{main:theorem2} $\vspace{1mm}$

a) We have the following information about the spectrum of $H^{++}_{[1,L]}$
restricted to $\Hil_{L,n}$:
$$
\lambda_{L,n}(1),\dots \lambda_{L,n}(L-n+1) 
  \in  [A(\Delta) - O(q^n),A(\Delta) + O(q^n)]\, ,
$$
and

b) $\lambda_{L,n}(L-n+2) \geq A(\Delta) + \gamma - O(n^{-1/4})$.

c) We have the following information about the eigenspace for the low-energy
states, $\lambda_{L,n}(1),\dots,\lambda_{L,n}(L-n+1)$:
$$
\|\Proj(\calK_{L,n}) - \Proj(\Hil^{L-n+1}_{L,n})\|
= O(q^{n/2})\, . 
$$
Equivalently
\begin{eqnarray*}
\sup_{0 \neq \psi \in \calK_{L,n}} 
  \left( \inf_{\psi' \in \Hil^{L-n+1}_{L,n}} 
  \frac{\|\psi - \psi'\|^2}{\|\psi\|^2} \right) = O(q^n)\, , \\
\sup_{0 \neq \psi' \in \Hil^{L-n+1}_{L,n}}
  \left( \inf_{\psi \in \calK_{L,n}}
  \frac{\|\psi - \psi'\|^2}{\|\psi'\|^2} \right) = O(q^n)\, .
\end{eqnarray*}
\end{theorem}
Figure \ref{fig:dropspec} illustrates the spectrum for a specific choice
of $L$ and $q$.
Note that Theorem \ref{main:theorem2} also implies that, for any sequence of
states with energies converging to $A(\Delta)$, we must have that the distances
of these states to the subspaces $\calK_{L,n}$ converges to zero.
The remainder of the paper is organized as follows.

Section \ref{sec:Ham} reviews some preliminary properties of the 
Hamiltonians that appear in the paper: a simple estimate for the gap
above the ground state of the XXZ Hamiltonian on an open chain without 
boundary terms, the spectral gap for the Hamiltonian with kink and
antikink boundary terms, and a preliminary lower bound for the energy
of a droplet state.

The proof of the main theorems is given in Sections 
\ref{Sect:Eval}, \ref{sec:polarized}, and \ref{Sec:ROP}. 
First, in Section \ref{Sect:Eval}, we calculate the energy of the 
proposed droplet states $\xi_{L,n}(x)$, defined in 
\eq{def_drop}. We also prove that these states are approximate
eigenstates. 

In Section \ref{sec:polarized}, we prove a basic estimate which 
shows that, given a state $\psi$ of the chain on $[1,L]$,
with energy $E$, there exists an interval $J\subset [1,L]$, of length 
$\vert J\vert$, where the state is fully polarized (i.e., {\em all up} 
or {\em all down} spins) with high probability. We obtain the following
lower bound for this probability:
$$
{\rm Prob}_\psi[\mbox{\small the spins in $J$ are all up or all down}]
\geq 1-\mbox{Constant}\times \vert J\vert\times\frac{E}{L}.
$$
The meaning of this bound is clear. For fixed energy $E$, as $L$ increases it
becomes  more and more likely that there exists an interval $J$, of given
length $\vert J\vert$, where the system is in the {\em all up} or {\em all down}
state. Of course, the location of the interval $J$ in $[1,L]$ depends on
$\psi$. The spectral gap of the model enters through the constant. An
estimate of this kind should be expected to hold for any ferromagnetic 
model with a gap, as the interaction encourages like spins to aggregate.

Section \ref{Sec:ROP} contains the most intricate part of the proof. We
implement the idea that the presence of an interval of all up or all down spins
in a state, allows one to decouple the action of  the Hamiltonians on the
subsystems to the left and the right of this  interval. If the spins in the
interval are {\em down}, the Hamiltonian decouples into a sum of a kink and an
antikink Hamiltonian, for which it is  known that there is spectral gap. If the
spins in the interval are {\em up}, we do not immediately obtain an estimate for
the gap, but  we can repeat the argument for the two decoupled subsystems.
If there are a sufficiently large number of down spins in the original 
system, this procedure must lead to an interval of {\em down} spins , and hence
an estimate for the gap, after a finite number of iterations. 

We will also prove, in Section \ref{Sec:Periodic-Infinite}, 
the analogous statements for rings and for the infinite chain with a large 
but finite number of down spins. 

Some calculations that are used in the proofs are collected in two appendices.

%%%%%%%%%%%%%%%%%%%%%%%%%%%%%%%%%%%%%%%%%%%%%%%%%%%%%%%%%%%%%%%%%%%%%%%%%%
%%%%%%%%%%%%%%%%%%%%%%%%%%%%%%%%%%%%%%%%%%%%%%%%%%%%%%%%%%%%%%%%%%%%%%%%%%
%%%%%%%%%%%%%%%%%%%%%%%%%%%%%%%%%%%%%%%%%%%%%%%%%%%%%%%%%%%%%%%%%%%%%%%%%%

\section{Properties of the XXZ Hamiltonians}
\label{sec:Ham}

In this section, we collect all the Hamiltonians that appear
in the paper, and describe some of their properties.
The first Hamiltonian we consider is
\begin{equation}
\label{xxzhamdef}
  \HXXZ_{[1,L]} = \sum_{x=1}^{L-1} \HXXZ_{x,x+1}\, 
\end{equation}
where
\begin{equation}
\label{XXZ:Ham2}
\HXXZ_{x,x+1} 
  = - \Delta^{-1} (\vec{S}_x \cdot \vec{S}_{x+1} - \frac{1}{4})
  - (1 - \Delta^{-1}) (S_x^{(3)} S_{x+1}^{(3)} - \frac{1}{4})\, .
\end{equation}
$\Delta > 1$ is the anisotropy parameter. 
Note that for $\Delta=1$ it is the isotropic
Heisenberg model, and for $\Delta = \infty$ it is the Ising model.

The diagonalization of $\HXXZ_{x,x+1}$, considered as an operator on
the four dimensional space $\Cx_x^2 \otimes \Cx_{x+1}^2$ is
\begin{equation}
\label{XXZ-nn:Diag}
\HXXZ_{x,x+1} : 
\qquad
\begin{array}{|c|c|}
  \rm{eigenvalue} & \rm{eigenvector} \\
  \hline 
  0 & \ket{\uparrow \uparrow},\, 
  \ket{\downarrow \downarrow} \\
  \frac{1}{2}(1 - \Delta^{-1})  & \frac{1}{\sqrt{2}} 
  (\ket{\uparrow \downarrow} + \ket{\downarrow \uparrow})\\
  \frac{1}{2}(1 + \Delta^{-1})  & \frac{1}{\sqrt{2}} 
  (\ket{\uparrow \downarrow} - \ket{\downarrow \uparrow})
\end{array}
\end{equation}
Let us define  
\begin{equation}
\label{pdef}
P^\sigma_{x,x+1} = \unity_1 \otimes \dots \otimes \unity_{x-1}
\otimes \ket{\sigma \sigma}\bra{\sigma \sigma} \otimes \unity_{x+2} \otimes
\dots \otimes \unity_L
\end{equation}
for $\sigma = \uparrow,\downarrow$,
and $P_{x,x+1} = P^\uparrow_{x,x+1} + P^\downarrow_{x,x+1}$.
Then, clearly,  
\begin{equation}
\label{localxxzgap} 
\HXXZ_{x,x+1} \geq \frac{1}{2}(1 - \Delta^{-1}) (\unity - P_{x,x+1})\, .
\end{equation}
\begin{lemma}
\label{XXZ-gap:Lemma}
The ground state energy for $\HXXZ_{[1,L]}$ is $0$,
and the ground state space is 
$\Span \{ \allup, \alldown \}$.
The following bounds hold
\begin{equation}
\label{XXZ-gap:bound}
  \HXXZ_{[1,L]} \geq \frac{1}{2}(1 - \Delta^{-1}) 
\Big(\unity - \Proj(\Span \{\allup,\alldown\})\Big)\, .
\end{equation}
\end{lemma}

\begin{proof}
The fact that $\allup$ and $\alldown$ are annihilated by $\HXXZ_{[1,L]}$
follows trivially from the fact that $\allup$ and $\alldown$ are annihilated
by each pairwise interaction $\HXXZ_{x,x+1}$.
So, in fact these states are frustration-free ground states.
Next,
$$
\HXXZ_{[1,L]}
  \geq \frac{1}{2}(1 - \Delta^{-1}) \sum_{x=1}^{L-1} (\unity - P_{x,x+1})\, ,
$$
by \eq{xxzhamdef} and \eq{localxxzgap}.
We observe that each $P_{x,x+1}$ is an orthogonal projection.
Moreover $P_{x,x+1}$ commutes with $P_{y,y+1}$ for every $x$ and $y$.
So
$$
\unity - \prod_{x=1}^{L-1} P_{x,x+1} 
  = \sum_{x=1}^{L-1} \left(\prod_{y=1}^{x-1} P_{y,y+1} \right) 
  (\unity - P_{x,x+1}) 
  \leq \sum_{x=1}^{L-1} (\unity - P_{x,x+1})\, .
$$
But $\prod_{x=1}^{L-1} P_{x,x+1} = \Proj(\Span\{\allup,\alldown\})$,
which proves \eq{XXZ-gap:bound}.
\end{proof}

All the other Hamiltonians we consider, namely $H^{\alpha \beta}_{[1,L]}$ 
for $\alpha,\beta=\pm 1,0$, defined in \eq{Hab}, are perturbations of 
$\HXXZ_{[1,L]}$ by boundary fields.
The Hamiltonian $H^{+-}_{[1,L]}$ is known as the kink Hamiltonian, 
and $H^{-+}_{[1,L]}$ is the antikink Hamiltonian.
These two models are distinguished 
because they each possess a quantum group symmetry, for the quantum group
$SU_q(2)$.
It should be mentioned that the representation of $SU_q(2)$ on $\Hil_L$
which commutes with $H^{+-}_{[1,L]}$ is different than the representation
which commutes with $H^{-+}_{[1,L]}$.
These Hamiltonians are also distinguished because, like $\HXXZ_{[1,L]}$,
they can be written as sums of nearest-neighbor interactions
and all their ground states are frustration-free.
We will give a formula, sufficient for our purposes,
for the ground states of $H^{+-}_{[1,L]}$ and $H^{-+}_{[1,L]}$, respectively.
First define the sectors of fixed total down-spins so that
$\Hil_{L,0} = \Span\{\allup\}$, and for $n=1,\dots,L$
$$
  \Hil_{L,n} = \Span \{ \left(\prod_{i=1}^n S_{x_i}^{-}\right) \allup : 
  1 \leq x_1 < x_2 < \dots < x_n \leq L \}\, .
$$
Thus, $S^3_{tot} {\rm Proj}(\Hil_{L,n}) 
= (\frac{L}{2} - n) {\rm Proj}(\Hil_{L,n})$.
Then $H^{+-}_{[1,L]}$ and $H^{-+}_{[1,L]}$ each have $L+1$ ground states,
one for each sector.
Let $\psi^{+-}_{[1,L]}(n)$ and $\psi^{-+}_{[1,L]}(n)$ be these ground states,
normalized as given in \eq{Intro:+-} and \eq{Intro:-+}. 
The spectral gap is known to exist
for each sector $\Hil_{L,n}$, $n=1,\dots,L-1$, and to be independent
of $n$.
Specifically, in \cite{KN1} the following was proved
\begin{proposition}
\label{kink-gap:Prop}
For the $SU_q(2)$ invariant Hamiltonian $H^{+-}_{[1,L]}$, $L \geq 2$, 
and $\Delta \geq 1$ one has
\begin{eqnarray*}
  \gamma_L &:=& \inf \left\{ \frac{\ip{\psi}{H^{+-}_{[1,L]} \psi}}
  {\ip{\psi}{\psi}}\, :\, 
  \psi \in \Hil_{L,n}\, , \psi \neq 0\, ,
  \ip{\psi}{\psi^{+-}_{[1,L]}}=0 \right\} \\
  &=& 1 - \Delta^{-1} \cos(\pi/L)\, .
\end{eqnarray*}
In particular
$$
  \gamma_L \geq 1 - \Delta^{-1},
$$
for all $L \geq 2$, and in addition the spectral gap above any of the
ground state representations of the GNS
Hamiltonian for the infinite chain is exactly $1 - \Delta^{-1}$.
\QED \quad
\end{proposition}

We will define $\gamma = 1 - \Delta^{-1}$ 
which is the greatest lower bound of all $\gamma_L$, and the 
spectral gap for the infinite chain.
A result identical with this one holds for the $H^{-+}_{[1,L]}$ spin chain,
which may be obtained using spin-flip or reflection symmetry.

There are important differences between the droplet Hamiltonian, 
$H^{++}_{[1,L]}$, and the kink Hamiltonian, which we briefly explain.
Since $H^{++}_{[1,L]}$ commutes with $S^3_{tot}$, it makes sense to
block diagonalize it with respect to the sectors $\Hil_{L,n}$,
$n=0,\dots,L$.
If we consider the spectrum of $H^{++}_{[1,L]}$ on the sector $\Hil_{L,n}$
for $L$ and $n$ both large, we will see that there are $L+1-n$ eigevalues
in a very small interval about $A(\Delta)$.
Then there is a gap above $A(\Delta)$ of width approximately $\gamma$,
with error at most $O(n^{-1/4})$,
which is free of any eigenvalues.
This is different from the case of the kink and antikink Hamiltonians where
the ground state in each sector is nondegenerate, with a uniform spectral
gap above. In our case, the ground state is non-degenerate only because
the translation invariance is broken in the finite systems. As $L\to\infty$,
the translation invariance is restored and the lowest eigenvalue in each sector
becomes infinitely degenerate. Therefore, as is done in Theorem 
\ref{main:theorem2}, it is natural to consider the spectral projection 
corresponding to the $L+1-n$ lowest eigenvalues as opposed to just the 
ground state space. 

Before beginning to prove the main theorem, we will observe some simple facts
about the droplet Hamiltonian.
First, the two site Hamiltonian $H^{++}_{x,x+1}$ restricted to 
$\Cx_x^2 \otimes \Cx_{x+1}^2$ is diagonalized as follows
\begin{equation}
\label{Hpp-2site:Diag}
H^{++}_{x,x+1} : 
\qquad
\begin{array}{|c|c|}
\rm{eigenvalue} & \rm{eigenvector} \\
  \hline 
-A(\Delta) & \ket{\uparrow \uparrow} \\
 \frac{1}{2}(1 - \Delta^{-1}) & \frac{1}{\sqrt{2}}
  (\ket{\uparrow \downarrow} + \ket{\downarrow \uparrow}) \\
  A(\Delta) & \ket{\downarrow \downarrow} \\
  \frac{1}{2}(1 + \Delta^{-1}) & \frac{1}{\sqrt{2}} 
  (\ket{\uparrow \downarrow} - \ket{\downarrow \uparrow})
\end{array}
\end{equation}
Note that it is \textit{not} true that $\Hpp_L$ is the sum of
$\Hpp_{x,x+1}$ for all nearest neighbor pairs $x,x+1 \in [1,L]$
as was the case for $\HXXZ_L$ and $H^{+-}_L$.
Instead the following identities are true:
\begin{eqnarray}
\label{useful1}
\Hpp_L &=& H^{+-}_{[1,x]} + H^{++}_{x,x+1} + H^{-+}_{[x+1,L]}\, ,\\
\label{useful2}
  &=& H^{+-}_{[1,x]} + H^{++}_{[x,L]}\, ,\\
\label{useful3}
  &=& H^{++}_{[1,x]} + H^{-+}_{[x,L]}\, ,
\end{eqnarray}
for $1\leq x\leq L-1$. These identities should be kept in mind
since they allow us to \textit{cut} the droplet spin chain at the
sites $x,x+1$.
This vague notion will be explained in detail in Section \ref{Sec:ROP}.
The diagonalization of $H^{--}_{x,x+1}$ is the same as the 
diagonalization of $H^{++}_{x,x+1}$ above, except that $\uparrow$
and $\downarrow$ are interchanged for each of the eigenvectors.

Now we state an obvious (but poor) preliminary lower bound for 
$\lambda_{L,n}(1)$.
\begin{proposition}
\label{APBound:Prop}
The ground state energy of $\Hpp_L$ on $\Hil_L$ is $-A(\Delta)$,
and the ground state space is $\Span\{\allup\}$.
Moreover, 
\begin{equation}
\label{hard:APBound}
\Rayleigh{\psi}{H^{++}_{[1,L]}} 
  \geq - A(\Delta) + \frac{1}{2}(1 - \Delta^{-1})
  \quad \textrm{for all nonzero}
  \  \psi \perp \allup\, .
\end{equation}
\end{proposition}

\begin{proof}
First, $H^{++}_{[1,L]} \geq -A(\Delta) \unity$ 
because $\HXXZ_{[1,L]} \geq 0$ and
$-A(\Delta) (S_1^{(3)} + S_L^{(3)}) \geq -A(\Delta) \unity$.
It is also clear that $H^{++}_{[1,L]} \allup = -A(\Delta) \allup$, and
$\Hpp_L \alldown = A(\Delta) \alldown$,
in agreement with \eq{hard:APBound}. 
Because $\allup$ and $\alldown$ are eigenvectors of the self-adjoint
operator $H^{++}_{[1,L]}$, all that remains is to check that
\eq{hard:APBound} holds on $\Span\{\allup,\alldown\}^\perp$.
But this is true by Lemma \ref{XXZ-gap:Lemma},
since $\Hpp_L \geq -A(\Delta) + \HXXZ_L$ and 
$\HXXZ_L \geq \frac{1}{2}(1 - \Delta^{-1})$
on $\Span\{\allup,\alldown\}^\perp$.
\end{proof}
We now begin the actual proof of the Theorems \ref{main:theorem}
and \ref{main:theorem2}.

%%%%%%%%%%%%%%%%%%%%%%%%%%%%%%%%%%%%%%%%%%%%%%%%%%%%%%%%%%
%%%%%%%% EVALUATION OF H++ ON DROPLET STATES     %%%%%%%%%
%%%%%%%%%%%%%%%%%%%%%%%%%%%%%%%%%%%%%%%%%%%%%%%%%%%%%%%%%%
\section{Evaluation of $H^{++}_{[1,L]}$ on droplet states.}
\label{Sect:Eval}

We begin by proving part (a) of Theorem \ref{main:theorem}.
This is straightforward because we have closed expressions for
each $\xi_{L,n}(x)$ and for $H^{++}_{[1,L]}$.
The heart of the proof is a number of computations which show that
$\xi_{L,n}(x)$ and $\xi_{L,n}(y)$ are approximately
orthogonal with respect to the inner product $\ip{*}{*}$ as well
as $\ip{*}{H^{++}_{[1,L]} *}$ and $\ip{*}{(H^{++}_{[1,L]})^2 *}$,
when $x \neq y$ and $n$ is large enough.
Specifically,
\begin{eqnarray}
\label{Refer:app:res1}
\frac{|\ip{\xi_{L,n}(x)}{\xi_{L,n}(y)}|}
  {\|\xi_{L,n}(x)\| \cdot \|\xi_{L,n}(y)\|}
  &\leq& \frac{q^{n|y-x|}}{\pn \infty} 
  \quad \textrm{for all} \quad x,y\, ;\\
\label{Refer:app:res2}
\frac{|\ip{\xi_{L,n}(x)}{H^{++}_{[1,L]} \xi_{L,n}(y)}|}
  {\|\xi_{L,n}(x)\| \cdot \|\xi_{L,n}(y)\|}
  &\leq& \frac{q^{n|y-x|}}{\pn \infty} 
  \quad \textrm{if} \quad x\neq y\, ;\\
\label{Refer:app:res3}
\frac{|\ip{\xi_{L,n}(x)}{(H^{++}_{[1,L]})^2 \xi_{L,n}(y)}|}
  {\|\xi_{L,n}(x)\| \cdot \|\xi_{L,n}(y)\|}
  &\leq& \frac{q^{n|y-x|}}{\pn \infty} 
  \quad \textrm{if} \quad |x-y| \geq 2\, .
\end{eqnarray}
Here $\pn \infty$ is a number arising in partition theory \cite{And},
$$
\pn \infty = \prod_{n=1}^\infty (1 - q^{2n})\, .
$$
(It is usually written as $(q^2;q^2)_\infty$.)
The important fact is that $\pn \infty \in (0,1]$ for
$q \in [0,1)$.

We need one more piece of information, which is that
\begin{equation}
\label{nec:detail}
\frac{\|(H^{++}_{[1,L]} - A(\Delta)) \xi_{L,n}(x)\|^2}
  {\|\xi_{L,n}(x)\|^2}
  \leq \frac{2 q^{2 \floor{n/2}}}{1 - q^{2 \floor{n/2}}}\, .
\end{equation}
To prove this, we refer to equation (6.7) of \cite{BCN}.
In that paper, it is proved that
$$
\frac{\|P^\downarrow_{L} \psi^{-+}_{[1,L]}(n)\|^2}{\|\psi^{-+}_{[1,L]}(n)\|^2} 
  < q^{2(L-n)} \frac{1-q^{2n}}{1- q^{2L}} 
  \leq \frac{q^{2(L-n)}}{1-q^{2(L-n)}}\, ,
$$
where 
$$
P^\sigma_{x} = \unity_1 \otimes \dots \otimes \unity_{x-1}
  \otimes \ket{\sigma}\bra{\sigma} \otimes \unity_{x+1} \otimes
  \dots \otimes \unity_L
$$
for $\sigma = \uparrow,\downarrow$.
Using spin-flip and reflection symmetry, we obtain
$$
\frac{\|P^\uparrow_{L} \psi^{+-}_{[1,L]}(n)\|^2}
  {\|\psi^{+-}_{[1,L]}(n)\|^2} 
  < \frac{q^{2n}}{1-q^{2n}}\, , \qquad
\frac{\|P^\uparrow_{1} \psi^{-+}_{[1,L]}(n)\|^2}
  {\|\psi^{-+}_{[1,L]}(n)\|^2} 
  < \frac{q^{2n}}{1-q^{2n}}\, .
$$
Since $\xi_{L,n}(x) = \psi^{+-}_{[1,x]}(\floor{n/2}) \otimes
\psi^{-+}_{[x+1,L]}(\ceil{n/2})$, we then have the bounds
\begin{equation}
\label{bcnsbounds}  
 \frac{\|P^\uparrow_{x} \xi_{L,n}(x)\|^2}{\|\xi_{L,n}(x)\|^2} 
  \leq \frac{q^{2\floor{n/2}}}{1 - q^{2\floor{n/2}}}\, ,\quad 
 \frac{\|P^\uparrow_{x+1} \xi_{L,n}(x)\|^2}{\|\xi_{L,n}(x)\|^2} 
  \leq \frac{q^{2\ceil{n/2}}}{1 - q^{2\ceil{n/2}}}\, .
\end{equation}
Now $H^{++}_{[1,L]} \xi_{L,n}(x) = H^{++}_{x,x+1} \xi_{L,n}(x)$, because
of the identity \eq{useful1}, and the fact that
$$
H^{+-}_{[1,x]} \xi_{L,n}(x) = H^{-+}_{[x+1,L]} \xi_{L,n}(x) = 0\, .
$$
By \eq{Hpp-2site:Diag}, we estimate
$$
0 \leq (H^{++}_{x,x+1} - A(\Delta))^2
  \leq  P^\uparrow_x + P^\uparrow_{x+1}\, ,
$$
which, together with \eq{bcnsbounds}, proves \eq{nec:detail}.

We are now poised to prove Theorem \ref{main:theorem} (a).
We state the argument, which is very simple, as a lemma.
It is useful to do it this way, because we will repeat the argument
twice more in the proofs of Theorems \ref{Thm:Periodic} and \ref{Thm:Infinite}.
\begin{lemma}
\label{Lem:OrthStates}
Let $\{f_n : n \in \Ir\}$ be a family of states, normalized so that
$\|f_n\| = 1$ for all $n$, but not necessarily orthogonal.
Suppose, however, that there are constants $C < \infty$ and $\epsilon < 1$
such that $|\ip{f_n}{f_m}| \leq C \epsilon^{|n-m|}$ for all $m,n$. 
If  $(1 + 2 C) \epsilon < 1$, then
\begin{equation}
\label{Lem:orth1}
\left\| \sum_{n\in\Ir} \Proj(f_n) 
  - \Proj(\Span(\{f_n : n\in \Ir\})) \right\|
  \leq \frac{2 C \epsilon}{1 - \epsilon}\, . 
\end{equation}
Suppose that $X$ is a self-adjoint operator such that for some $r<\infty$
we have $\|X f_n\| \leq r$ for all $n$, 
and for some $C'<\infty$, $N \in \Nl$ we have
$|\ip{X f_n}{X f_m}| \leq C' \epsilon^{|n-m|}$ whenever $|n-m| \geq N$.
Then
\begin{equation}
\label{Lem:orth2}
\left\| X \cdot \Proj(\Span(\{f_n : n \in \Ir\})) \right\| 
  \leq \left[ \frac{(2N-1) r^2 + \frac{2 C' \epsilon^N}{1-\epsilon}}
  {1 - \frac{2 C \epsilon}{1 - \epsilon}} \right]^{1/2} \, .
\end{equation}
The same results hold if $\{f_n\}$ is a finite family, 
in which case the bounds are even smaller.
\end{lemma}

\begin{proof}
Define $F = \sum_{n=-\infty}^\infty \ket{f_n}\bra{f_n}$.
Define $E$ an infinite matrix such that $E_{mn} = \ip{f_m}{f_n}$.
Let $\{e_n : n \in \Ir\}$ be an orthonormal family in any Hilbert
space, and let $A = \sum_n \ket{f_n} \bra{e_n}$.
Then $E = A^* A$ and $F = A A^*$. 
For simplicity let $\mathcal{F}=\textrm{cl}(\Span(\{f_n : n \in \Ir\}))$,
and let $\mathcal{E}=\textrm{cl}(\Span(\{e_n : n \in \Ir\}))$.
We consider $A : \mathcal{E} \to \mathcal{F}$.
Then we calculate
$$
\|A^* A - \unity_{\mathcal{E}}\|
  \leq \sup_m \sum_{n \atop n\neq m} |E_{mn}|
  \leq \frac{2 C \epsilon}{1 - \epsilon}\, .
$$
Since $2 C \epsilon < 1 - \epsilon$,
this shows that $A$ is bounded and $A^* A$ is invertible.
Under the invertibility condition, it is true that $A A^*$ is also
invertible on $\mathcal{F}$, and considering this as its domain,
$\sigma(A A^*) = \sigma(A^* A)$.
If we let $E$ and $F$ operate on proper superspaces of
$\mathcal{E}$ and $\mathcal{F}$,
then they will be identically zero on the orthogonal complements.
But it is still true that
$$
\sigma(E) \setminus \{0\} 
  = \sigma(A A^*) 
  = \sigma(A^* A) 
  = \sigma(F) \setminus \{0\}\, .
$$
In particular, if we let $P_{\mathcal{F}}$ be the orthogonal projection
onto $\mathcal{F}$, 
then
$$
\|F - P_{\mathcal{F}}\| = \|A^*A - \unity_{\mathcal{E}}\|
\leq \frac{2 C \epsilon}{1 - \epsilon}\, .
$$
This proves \eq{Lem:orth1}.

To prove the second part, let $\psi = \sum_n \alpha_n f_n$
be a state in $\mathcal{F}$.
Let $\phi = \sum_n \alpha_n e_n$.
Then 
\begin{equation}
\label{Proof:orth1}
\|\psi\|^2 = \ip{\phi}{A^*A\phi} 
  \geq (1 - \frac{2C\epsilon}{1-\epsilon}) \sum_n |\alpha_n|^2\, .
\end{equation}
We calculate 
\begin{eqnarray*}
\|X \psi\| 
  = \sum_{m,n} \overline{\alpha}_m \alpha_n \ip{X f_m}{X f_n} 
  \leq \sum_n |\alpha_n|^2 \cdot \sup_m 
  \sum_{n} |\ip{X f_m}{X f_n}|\, .
\end{eqnarray*}
Breaking the sum into two pieces yields,
for any $m \in \Ir$,
\begin{eqnarray*}
\sum_n |\ip{X f_m}{X f_n}|
  &\leq& \sum_{n \atop |m-n| < N} |\ip{X f_m}{X f_n}|
  + \sum_{n \atop |m-n| \geq N} |\ip{X f_m}{X f_n}| \\
  &\leq& (2N-1) r^2 + \frac{2 C' \epsilon^N}{1-\epsilon}\, .
\end{eqnarray*}
So, using \eq{Proof:orth1}, we have
$$
\frac{\|X \psi\|^2}{\|\psi\|^2}
  \leq \frac{(2N-1) r^2 + \frac{2 C' \epsilon^N}{1 - \epsilon}}
  {1 - \frac{2 C \epsilon}{1 - \epsilon}} 
$$
for any nonzero $\psi \in \mathcal{F}$.
This proves \eq{Lem:orth2}.
\end{proof}

Now to prove Theorem \ref{main:theorem}(a), we note that
the hypotheses of the lemma are met.
Namely, take $f_x = \xi_{L,n}(x)$.
By \eq{Refer:app:res1}, we have $|\ip{f_x}{f_y}| \leq C \epsilon^{|x-y|}$,
where $C = f_q(\infty)^{-1}$ and $\epsilon = q^n$.
We set $X = H^{++}_{[1,L]} - A(\Delta)$.
Then by \eq{Refer:app:res1}, \eq{Refer:app:res2} and \eq{Refer:app:res3},
we have $\ip{X f_x}{X f_y} \leq C' \epsilon^{|x-y|}$, for $|x-y| \geq 2$,
where $C' = 4/\pn{\infty}$. (Since $A(\Delta) \leq 1$,
$1 + 2 A(\Delta) + A(\Delta)^2 \leq 4$.)
By \eq{nec:detail}, we have $\|X \xi_x\| \leq r$ for all $x$, where
$r^2 = 2 q^{2 \floor{n/2}}/(1 - q^{2 \floor{n/2}})$.
Therefore, by Lemma \ref{Lem:OrthStates}, and some trivial estimations
\begin{equation}
\label{Eval:result1}
\|(H^{++}_{[1,L]} - A(\Delta))\cdot \Proj(\calK_{L,n})\| 
  \leq  \frac{2 \sqrt{2} q^{\floor{n/2}}}
  {\sqrt{(1 - 3 q^{2 \floor{n/2}}) f_q(\infty)}}\, .
\end{equation}
The lemma also gives us the following result
\begin{equation}
\label{Eval:result2}
\|\Proj(\calK_{L,n}) 
  - \sum_{x=\floor{n/2}}^{L-\ceil{n/2}} \Proj(\xi_{L,n}(x))\|
  \leq \frac{2 q^n}{(1-q^n) f_q(\infty)}\, .
\end{equation}
This will prove useful in Section \ref{Sec:ROP}, 
because it is a precise statement of  
just how orthogonal our proposed states $\xi_{L,n}(x)$ are
to each other.

%%%%%%%%%%%%%%%%%%%%%%%%%%%%%%%%%%%%%%%%%%%%%%%%%%%%%%%%%%
%%%%%%%%%   INTERVALS OF HOMOGENEOUS SPIN   %%%%%%%%%%%%%%
%%%%%%%%%%%%%%%%%%%%%%%%%%%%%%%%%%%%%%%%%%%%%%%%%%%%%%%%%%

\section{Existence of fully polarized intervals}
\label{sec:polarized}

We know that the ground states of the kink Hamiltonian exhibit a 
localized interface such that to the left of the interface 
nearly all spins are observed in the $\downarrow$ state,
and to the right nearly all spins are observed in the
$\uparrow$ state. 
The interface has a thickness due to quantum fluctuations.
A similar phenomenon occurs with the antikink Hamiltonian but with left
and right reversed or alternatively with $\uparrow$ and $\downarrow$ 
reversed.
We might hope that the ground state of the droplet Hamiltonian will
also contain an interval (or several intervals) 
with nearly all $\uparrow$- or all $\downarrow$-spins.
This is the case, and we prove it next.

\begin{definition}
For any finite interval $J \subset  \Ir$
define the orthogonal projections
\begin{eqnarray*}
P^\uparrow_J  &=& \ket{\uparrow \dots \uparrow} 
  \bra{\uparrow \dots \uparrow}_J \otimes \unity_{I \setminus J}\, ,\\
P^\downarrow_J  &=& \ket{\downarrow \dots \downarrow} 
  \bra{\downarrow \dots \downarrow}_J \otimes \unity_{I \setminus J}\, ,\\
P_J &=& P^\uparrow_J + P^\downarrow_J\, .  
\end{eqnarray*}

We also define for any operator $X$ and any nonzero state $\psi$,
the Rayleigh quotient
$$
\rho(\psi,X) = \Rayleigh{\psi}{X}\, .
$$
\end{definition}

\begin{proposition}
\label{IHS:Prop}
Suppose $\psi \in \Hil_L$ is a nonzero state, and let
$$
E = \rho(\psi,\HXXZ_L)\, .
$$
Given $l < L$, there is a subinterval $J = [a,a+l-1] \subset [1,L]$
satisfying the bound
\begin{equation}
\label{proj:prox}
  \frac{\|P_J \psi\|^2}{\|\psi\|^2}
  \geq 1 - \frac{2 E}{\gamma \lfloor{L/l}\rfloor}\, .
\end{equation}
Moreover denoting 
$$
\epsilon := \frac{2E}{\gamma \lfloor{L/l}\rfloor}\, ,
$$ 
then as long as $\epsilon < 1$, we have the following bound
\begin{equation}
\label{ihs:en:comp}
\rho(P_J \psi,\HXXZ_{[1,L]})
  \leq \frac{E}{1 - \epsilon} + 
  2 \Delta^{-1} \sqrt{\frac{\epsilon}{1-\epsilon}}\, .
\end{equation}
\end{proposition}

\begin{proof}
Partition $[1,L]$ into $r=\lfloor{L/l}\rfloor$ intervals $J_1,\dots,J_r$
each of length $\geq l$.
If $J_i = [a_i,a_{i+1}-1]$ then 
$$
\HXXZ_L = \sum_{i=1}^r \HXXZ_{J_i} + \sum_{i=2}^r \HXXZ_{a_i-1,a_i}\,
  \geq \sum_{i=1}^r \HXXZ_{J_i}\, .
$$ 
By Lemma \ref{XXZ-gap:Lemma}, 
$$
\rho(\psi,\HXXZ_{J_i}) 
  \geq \frac{\gamma}{2} (1 - \rho(\psi,P_{J_i}))\, .
$$
So
\begin{eqnarray*}
E \geq \frac{\gamma}{2} \sum_{i=1}^r (1 - \rho(\psi,P_{J_i}))
  \geq r \frac{\gamma}{2} \min_i (1 - \rho(\psi,P_{J_i}))\, .
\end{eqnarray*}
In other words,
$$
\rho(\psi,P_{J_i}) \geq 1 - \frac{2 E}{\gamma r}\, ,
$$
for some $i$.
Since $[a_i,a_i+l] \subset J_i$, $P_{J_i} \leq P_{[a_i,a_i+l+1]}$.
Let $J = [a_i,a_i+l-1]$, then \eq{proj:prox} holds.

Note that for any orthogonal projection $P$ and any operator $H$ we have 
the decomposition
$$
H = P H P + (1-P) H (1-P) + [P,[P,H]]\, .
$$
If $H$ is nonnegative, then $(1-P)H(1-P)$ is as well.
Hence
$$
P H P  \leq H  - [P,[P,H]]\, .
$$
On the other hand, it is obvious that
$$
P [P,[P,H]] P = (1-P) [P,[P,H]] (1-P) = 0\, ,
$$
which implies
$$
\rho(\psi,P H P) \leq \rho(\psi,H) 
  + 2 \|[P,[P,H]]\| 
  \frac{\|P \psi\|\, \|(1-P) \psi\|}{\|\psi\|^2}
$$
for any nonzero $\psi$.

Moreover,
\begin{equation}
\label{ray:ineq}
\rho(P \psi,H) = \frac{\rho(\psi,P H P)}{\rho(\psi,P)}
  \leq \frac{\rho(\psi,H)}{\rho(\psi,P)} 
  + 2 \|[P,[P,H]]\| \sqrt{\frac{\rho(\psi,1-P)}{\rho(\psi,P)}}\, .
\end{equation}
In our particular case, where
$H = \HXXZ_L$ and $P = P_J$, 
\eq{ray:ineq} and \eq{proj:prox} imply
\begin{equation}
\label{ihs:en:ineq}
\rho(P_J \psi,\HXXZ_L) 
  \leq \frac{E}{1-\epsilon} 
  + 2 \|[P_J,[P_J,\HXXZ_L]]\| \sqrt{\frac{\epsilon}{1-\epsilon}}\, .
\end{equation}
All that remains is to calculate $\|[P_J,[P_J,\HXXZ_{[1,L]}]]\|$.

Notice that 
$$
[P_J,[P_J,\HXXZ_{[1,L]}]] 
  = \sum_{x \in [1,L-1] \atop \alpha,\beta \in \{\uparrow,\downarrow\}}
  [P^\alpha_J,[P^\beta_J,\HXXZ_{x,x+1}]]\, ,
$$
and that $\HXXZ_{x,x+1}$
commutes with $P^\beta_J$ for all $x,x+1$ except $a-1,a$
and $b,b+1$. (We define $b=a+l-1$.)
Straightforward computations yield
$$
[P^{\beta}_J,\HXXZ_{a-1,a}] 
  = - \frac{1}{2 \Delta} \unity_{[1,a-2]} 
  \otimes (\ket{\beta \beta'} \bra{\beta' \beta} 
  - \ket{\beta' \beta} \bra{\beta \beta'}) \otimes P^{\beta}_{[a+1,b]}
  \otimes \unity_{[b+1,L]} 
$$
and
$$
[P^{\beta}_J,\HXXZ_{b,b+1}]
  = - \frac{1}{2 \Delta} \unity_{[1,a-1]} 
  \otimes P^\beta_{[a,b-1]} 
  \otimes (\ket{\beta \beta'} \bra{\beta' \beta} 
  - \ket{\beta' \beta} \bra{\beta \beta'})
  \otimes \unity_{[b+2,L]}\, ,
$$
where $\uparrow' = \downarrow$ and $\downarrow' = \uparrow$.
It is easy to deduce that $[P^\alpha_J,[P^\beta_J,\HXXZ_L]]$
is zero unless $\alpha=\beta$.
($[P^\beta_J,\HXXZ_{a-1,a}]$ has a tensor factor $P^\beta_{[a+1,b]}$
and $P^\alpha_J$ has a tensor factor $P^\alpha_{[a+1,b]}$, which
implies $[P^\alpha_J,[P^\beta_J,\HXXZ_{a-1,a}]]$ is zero unless
$\alpha = \beta$.
The term $[P^\alpha_J,[P^\beta_J,\HXXZ_{b,b+1}]]$ is treated similarly.)
Another straightforward computation yields
$$
[P^\beta_J,[P^{\beta}_J,\HXXZ_{a-1,a}]]
  = - \frac{1}{2 \Delta} \unity_{[1,a-2]} 
  \otimes (\ket{\beta \beta'} \bra{\beta' \beta} 
  + \ket{\beta' \beta} \bra{\beta \beta'}) \otimes P^{\beta}_{[a+1,b]}
  \otimes \unity_{[b+1,L]} 
$$
and
$$
[P^\beta_J,[P^{\beta}_J,\HXXZ_{b,b+1}]]
  = - \frac{1}{2 \Delta} \unity_{[1,a-1]} \otimes P^\beta_{[a,b-1]} 
  \otimes (\ket{\beta \beta'} \bra{\beta' \beta} 
  + \ket{\beta' \beta} \bra{\beta \beta'})
  \otimes \unity_{[b+2,L]}\, .
$$
So
$$
\begin{array}{l}
\displaystyle
[P_J,[P_J,\HXXZ_L]]
  = -\frac{1}{2 \Delta} \Big(
  \unity_{[1,a-2]} 
  \otimes A_{a-1,a}
  \otimes P_{[a+1,b]}
  \otimes \unity_{[b+1,L]} \\
\hspace{125pt} \displaystyle
  + \unity_{[1,a-1]} \otimes P_{[a,b-1]} 
  \otimes A_{b,b+1}
  \otimes \unity_{[b+2,L]} \Big)\, ,
\end{array}
$$
where
$A =  \ket{\uparrow \downarrow} \bra{\downarrow \uparrow} 
  + \ket{\downarrow \uparrow} \bra{\uparrow \downarrow}$.
In particular $\|A\| = 1$, so that
$$
\|\unity_{[1,a-2]} 
  \otimes A_{a-1,a}
  \otimes P_{[a+1,b]}
  \otimes \unity_{[b+1,L]}\|
  = 1\, ,
$$
and 
$$
\|\unity_{[1,a-1]} \otimes P_{[a,b-1]} 
  \otimes A_{b,b+1}
  \otimes \unity_{[b+2,L]} \| = 1\, .
$$
Thus
$\|[P_J,[P_J,\HXXZ_L]]\| \leq \Delta^{-1}$,
which along with \eq{ihs:en:ineq} proves \eq{ihs:en:comp}.
\end{proof}

%%%%%%%%%%%%%%%%%%%%%%%%%%%%%%%%%%%%%%%%%%%%%%%%%%%%%%%%%%
%%%%%%%%%%%%%%%%%   IHS COROLLARY   %%%%%%%%%%%%%%%%%%%%%%
%%%%%%%%%%%%%%%%%%%%%%%%%%%%%%%%%%%%%%%%%%%%%%%%%%%%%%%%%%

In the following corollary, we show that essentially the 
same results hold for any bounded perturbation of $\HXXZ_{[1,L]}$.

\begin{corollary}
\label{IHS:Cor}
\label{IHS:Prop2}
Suppose $H_L$ is a bounded operator on $\Hil_L$ with
$$
M = \|H_L - \HXXZ_{[1,L]}\|\, .
$$ 
Let $E<\infty$ and $\psi \in \Hil_L$ be a nonzero state with
$$
\rho(\psi,H_L) \leq E\, .
$$
Given any subinterval $K \subset [1,L]$ and $l < |K|$, there is a 
sub-subinterval $J \subset K$ of length $l$, satisfying the bound
\begin{equation}
\label{cor:proj:prox}
  \|\psi - P_J \psi\|^2
  \leq \epsilon \|\psi\|^2\, ,
\end{equation}
where
$$
\epsilon = \frac{2(E+M)}{\gamma \floor{|K|/|J|}}\, .
$$
This statement is nonvacuous when $\epsilon < 1$.
Also under the assumption that $\epsilon<1$, we have the bound
\begin{equation}
\label{cor:en:comp}
\ip{\psi}{H_L \psi}
  \geq \ip{P_J \psi}{H_L P_J \psi} -
 \left(M \epsilon
  + 2 (\Delta^{-1} + 2 M) \sqrt{\epsilon(1-\epsilon)}\right)\, .
\end{equation}
\end{corollary}

%%%%%%%%%%%%%%%%%%%%%%%%%%%%%%%%%%%%%%%%%%%%%%%%%%%%%%%%%%

\begin{proof}
Since $\|H_L - \HXXZ_{[1,L]}\| = M$, it is clear that
$$
\rho(\psi,\HXXZ_K) \leq \rho(\psi,\HXXZ_{[1,L]}) \leq E + M\, .
$$
So Proposition \ref{IHS:Prop} implies \eq{cor:proj:prox}.
To prove \eq{cor:en:comp} notice that for any operator $H$,
any orthogonal projection $P$, and any nonnegative operator $\tilde H$,
\begin{eqnarray*}
H - PHP 
  &=& (1-P)H(1-P) + [P,[P,H]] \\
  &=& (1-P)\tilde{H}(1-P) + (1-P)(H - \tilde{H})(1-P) \\
  && \qquad + [P,[P,\tilde{H}]] + [P,[P,H-\tilde{H}]] \\
  &\geq& (1-P)(H - \tilde{H})(1-P) + [P,[P,\tilde{H}]] \\
  && \qquad + [P,[P,H-\tilde{H}]]\, .
\end{eqnarray*}
So, for any nonzero $\psi$,
\begin{eqnarray*}
&& \rho(\psi,H - PHP)
  \geq - \|H - \tilde{H}\| \rho(\psi,1-P)  \\
  && \hspace{50pt}
  - 2 (\|[P,[P,\tilde{H}]]\| + 2 \|H - \tilde{H}\|) 
  \rho(\psi,P)^{1/2} \rho(\psi,1-P)^{1/2}\, .
\end{eqnarray*}
Setting $H = H_L$, $\tilde{H} = \HXXZ_L$ and $P = P_J$ we have
$$
\rho(\psi,H_L) - \rho(\psi,P_J H_L P_J)
  \geq - M \epsilon - 2 (\Delta^{-1} + 2 M) \sqrt{\epsilon (1-\epsilon)}\, .
$$
Since
$$
\rho(P_J \psi,H_L) = \frac{\rho(\psi,P_J H_L P_J)}{\rho(\psi,P_J)}
	\leq \frac{\rho(\psi,P_J H_L P_J)}{1-\epsilon},
$$
the corollary is proved.
\end{proof}

%%%%%%%%%%%%%%%%%%%%%%%%%%%%%%%%%%%%%%%%%%%%%%%%%%%%%%%%%%
%%%%%%%%%%%%%%% REMAINDER OF PROOF %%%%%%%%%%%%%%%%%%%%%%%
%%%%%%%%%%%%%%%%%%%%%%%%%%%%%%%%%%%%%%%%%%%%%%%%%%%%%%%%%%

\section{Remainder of the proof}
\label{Sec:ROP}

We will now prove Theorem \ref{main:theorem}(b).
Let us henceforth denote $\Proj(\Span\{\phi\})$ simply by
$\Proj(\phi)$ for any nonzero state $\phi$.
We observe by \eq{Eval:result2} that there are constants $C_0(q)$
and $N_0(q)$, such that
$$
\|\Proj(\calK_{L,n}) 
  - \sum_{x=\floor{n/2}}^{L-\ceil{n/2}} \Proj(\xi_{L}(x,n))\|
  \leq C_0(q) q^n\, .
$$
whenever $n \geq N_0(q)$.
By \eq{Eval:result2}, 
$N_0(q) = 1$ and $C_0(q) = (1-q)^{-1} \pn \infty ^{-1}$. 
Suppose we exhibit a sequence
$\epsilon_n$, with $\lim_{n \to \infty} \epsilon_n = 0$, such that
\begin{equation}
\begin{array}{l}
\displaystyle H^{++}_{[1,L]} \Proj(\Hil_{L,n}) \geq \vspace{1mm}\\
\quad \displaystyle
(A(\Delta) - \epsilon_n) \Proj(\Hil_{L,n})
  + \gamma [\Proj(\Hil_{L,n}) 
- \sum_{x=\floor{n/2}}^{L - \ceil{n/2}} \Proj(\xi_{L}(x,n))]\, .
\label{ROP:piece1}
\end{array}
\end{equation}
We know, by Theorem \ref{main:theorem}(a), that 
$(H^{++}_{[1,L]} - A(\Delta)) \Proj(\calK_{L,n})$ is bounded above
and below by $\pm C q^n \unity$.
Then we would know
\begin{eqnarray*}
&&H^{++}_{[1,L]} \Proj(\Hil_{L,n}) 
  \geq (A(\Delta) - 2 C q^n) \Proj(\Hil_{L,n}) + \\
&&\qquad \qquad  (\gamma - \epsilon_n) (\Proj(\Hil_{L,n}) - 
\Proj(\calK_{L,n}))\, .
\end{eqnarray*}
So to prove Theorem \ref{main:theorem}(b), it suffices to 
verify that there is a sequence $\epsilon_n$ satisfying 
\eq{ROP:piece1}.

We will prove this fact in this section.
We find it convenient to consider an arbitrary gap $\lambda$,
$0 \leq \lambda<\gamma$.
Define $\epsilon_\lambda(L,n)$ to be the smallest nonnegative
number such that 
$$
\ip{\psi}{H^{++}_{[1,L]} \psi}
  \geq (A(\Delta) - \epsilon_\lambda(L,n)) \|\psi\|^2
  + \lambda \ip{\psi}{[\unity - 
  \sum_{x=\floor{n/2}}^{L - \ceil{n/2}} \Proj(\xi_{L}(x,n))] \psi}
$$
holds for all $\psi \in \Hil_{L,n}$.
We also define 
\begin{eqnarray*}
\epsilon_\lambda'(L,n) &=& \max_{L' \atop n \leq L' \leq L} 
\epsilon_\lambda(L',n) \\
\epsilon_\lambda'(\infty,n) &=& \lim_{L \to \infty} 
\epsilon_\lambda'(L,n) \\
\epsilon_\lambda''(n) &=& \sup_{n' \atop n' \geq n} 
\epsilon_\lambda'(\infty,n')
\end{eqnarray*}
If we can prove that for every $\lambda<\gamma$,
$\lim_{n \to \infty} \epsilon_\lambda''(n) = 0$,
then we will have proved Theorem \ref{main:theorem}(b).

Given $0\leq q<1$, define
$$
N_1(q) = \left(\frac{5-4q+\sqrt{(6-5q)(4-3q)}}{1-q}\right)^2\, .
$$
Suppose $n > N_1(q)$ and $L \geq n$.
(The requirement that $n > N_1(q)$ allows us to apply Corollary \ref{IHS:Cor}
effectively, i.e.\ with $\epsilon < 1$.)
Define an interval $K = [\ceil{\frac{1}{4}L},\floor{\frac{3}{4}L}]$,
and suppose $\psi \in \Hil_{L,n}$ is a nonzero state with
$\rho(\psi,H^{++}_{[1,L]}) \leq A(\Delta) + \gamma$.
Then by Corollary \ref{IHS:Cor} and the requirement that $n>N_1(q)$, 
we can find an interval $J \subset K$ such that $|J| = \floor{L^{1/2}}$,
\begin{equation}
\label{ROP:corres1}
\|\psi - P_J \psi\|^2 \leq C_1(q) L^{-1/2} \|\psi\|^2\, ,
\end{equation}
and
\begin{equation}
\label{ROP:corres2}
\ip{\psi}{H^{++}_{[1,L]} \psi} 
  \geq \ip{P_J \psi}{H^{++}_{[1,L]} P_J \psi}
  - C_2(q) L^{-1/4} \|\psi\|^2\, ,
\end{equation}
where
$$
\begin{array}{l}
\displaystyle 
C_1(q) = \frac{8}{1-q} (1 - 2 n_1(q)^{-1/2} - n_1(q)^{-1})^{-1}\, ,
\vspace{2mm}\\
\displaystyle
C_2(q) = \frac{(1+3q)(3-q)}{2(1+q^2)} C_1(q)^{1/2}\, .
\end{array}
$$
Let $J = [a,b]$.

We need to extend our definition of $\Hil_{L,n}$ in the following way.
For integers $s \leq t$, let 
$$
\Hil_{[s,t]} = \Cx_s^2 \otimes \Cx_{s+1}^2 \otimes \cdots \otimes\Cx_t^2.
$$
For $0 \leq r \leq s-t+1$, let
$$
\Hil_{[s,t],r} = \Span\{ \left(\prod_{i=1}^r S_{x_i}^-\right)
  \ket{\uparrow \dots \uparrow}_{[s,t]} :
  s \leq x_1 < x_2 < \dots <x_r \leq t \}\, .
$$
So $\Hil_L = \Hil_{[1,L]}$ in the new notation,
and $\Hil_{L,n} = \Hil_{[1,L],n}$.
We are free to decompose
$$
\psi = \sum_{n_1,n_2,n_3} \psi(n_1,n_2,n_3) 
$$
where $\psi(n_1,n_2,n_3) \in \Hil_{[1,a-1],n_1} \otimes
\Hil_{[a,b],n_2} \otimes \Hil_{[b+1,L],n_3}$.
The condition that $\psi \in \Hil_{L,n}$ implies
$\psi(n_1,n_2,n_3) \neq 0$ only if
$(n_1,n_2,n_3) \in [0,a-1] \times [0,b-a+1] \times[0,L-b]$,
and $n_1+n_2+n_3 = n$.
Also, since the range of $P_J$ is precisely the direct sum of all those 
triples $\Hil_{[1,a-1],n_1} \otimes \Hil_{[a,b],n_2} \otimes \Hil_{[b+1,L],n_3}$
such that $n_2\in\{0,|J|\}$, we can restrict attention to 
those states $\psi(n_1,n_2,n_3)$ satisfying the same condition.
Therefore, let $\psi^\uparrow(j) = \psi(j,0,n-j)$,
and $\psi^\downarrow(j) = \psi(j,|J|,n-j-|J|)$.
Then $\psi^\uparrow(j)$ lies in the range of $P^\uparrow_J$
and $\psi^\downarrow(j)$ lies in the range of $P^\downarrow_J$,
and 
$$
P_J \psi = \sum_{j = 0}^{n} \psi^\uparrow(j)
    + \sum_{j=0}^{n-|J|} \psi^\downarrow(j)\, .
$$
Let $Q(n_1,n_2,n_3) = \Proj(\Hil_{[1,a-1]}^{n_1} \otimes
\Hil_{[a,b]}^{n_2} \otimes \Hil_{[b+1,L]}^{n_3})$.
Then it is easy to see that
$$
Q(n_1,n_2,n_3) \, H^{++}_{[1,L]} \, Q(m_1,m_2,m_3) = 0
$$ 
except when
$(n_1-m_1,n_2-m_2,n_3-m_3)$ equals $(\pm 1,\mp 1,0)$
or $(0,\pm 1,\mp 1)$.
But if $n_2,m_2 \in \{0,|J|\}$
(and $|J| > 1$), then the condition of the previous line can never
be met.
Therefore
\begin{equation}
\label{rop:1}
\ip{P_J \psi}{H^{++}_{[1,L]} P_j \psi}
  = \sum_{j=0}^n 
  \ip{\psi^\uparrow(j)}{H^{++}_{[1,L]} \psi^\uparrow(j)}
  + \sum_{j = 0}^{n-|J|}
  \ip{\psi^\downarrow(j)}{H^{++}_{[1,L]} \psi^\downarrow(j)}\, ,
\end{equation}
just as
\begin{equation}
\label{rop1a}
\|P_J \psi\|^2  
  = \sum_{j=0}^n \|\psi^\uparrow(j)\|^2
  + \sum_{j = 0}^{n-|J|} \|\psi^\downarrow(j)\|^2\, .
\end{equation}
We will next bound each of the terms on the right hand side
of \eq{rop:1}.

Let $x = a + \floor{|J|/2} = \floor{(a+b+1)/2}$.
Since $x,x+1 \in J$, consulting \eq{Hpp-2site:Diag}, we have
$$
H^{++}_{x,x+1} \psi^\downarrow(j) 
  = A(\Delta) \psi^\downarrow(j)\, .
$$
Then, by \eq{useful1}, it is clear
$$
\begin{array}{rcl}
\displaystyle
  \ip{\psi^\downarrow(j)}{H^{++}_{[1,L]} \psi^\downarrow(j)} 
  &\geq& A(\Delta) \|\psi^\downarrow(j)\|^2 \vspace{2mm}\\
\displaystyle  &&\quad + \ip{\psi^\downarrow(j)}
  {(H^{+-}_{[1,x]}+H^{-+}_{[x+1,L]}) \psi^\downarrow(j)} \, .
\end{array}
$$
By Proposition \ref{kink-gap:Prop}
$$
\begin{array}{l}
\displaystyle
\ip{\psi^\downarrow(j)}
  {(H^{+-}_{[1,x]}+H^{-+}_{[x+1,L]}) \psi^\downarrow(j)} 
  \geq \gamma \bra{\psi^\downarrow(j)} \vspace{2mm}\\
\hspace{1cm} \displaystyle
  \left(\unity - \Proj(\psi^{+-}_{[1,x]}(j') \otimes
  \psi^{-+}_{[x+1,L]}(n-j'))\right) \ket{\psi^\downarrow(n_1)}
\end{array}
$$
where $j' = j + \floor{|J|/2} + 1$.
Also, defining $\tilde{x}_j = a-1+\floor{n/2}-j$,
we know by \eq{App:result1}
$$
\|\Proj(\psi^{+-}_{[1,x]}(j') \otimes
  \psi^{-+}_{[x+1,L]}(n-j')) - \Proj(\xi_{L,n}(\tilde{x}_j)\|
  \leq C_3(q) q^{|J|/2} \, ,
$$
where
$C_3(q) = 4 (1 - q^2)^{-1/2}$.
Therefore,
\begin{equation}
\label{rop:2}
\begin{array}{l}
\displaystyle
\ip{\psi^\downarrow(n_1)}{H^{++}_{[1,L]} \psi^\downarrow(j)}
  \geq (A(\Delta) - C_3 q^{|J|/2}) \|\psi^\downarrow(j)\|^2 
\vspace{2mm}\\
\hspace{3cm} \displaystyle
+ \gamma \ip{\psi^\downarrow(j)}
  {\Big(\unity - \Proj(\xi_{L,n}(\tilde{x}_j))\Big)\psi^\downarrow(j)}\, .
\end{array}
\end{equation}

Next, we bound
$\ip{\psi^\uparrow(j)}{H^{++}_{[1,L]} \psi^\uparrow(j)}$
in the case that $1 \leq j \leq \floor{n/2}$.
The case $\floor{n/2} \leq j \leq n-1$, will be the same by symmetry.
Referring to \eq{useful2},
$$
\ip{\psi^\uparrow(j)}{H^{++}_{[1,L]} \psi^\uparrow(j)}
  = \ip{\psi^\uparrow(j)}{H^{+-}_{[1,x]} \psi^\uparrow(j)} 
  + \ip{\psi^\uparrow(j)}{H^{++}_{[x,L]} \psi^\uparrow(j)}\, .
$$
Now, since $\psi^\uparrow(j) \in \Hil_{[1,x-1],j} \otimes 
\Hil_{[x,L],n-j}$, we may bound
$$
\ip{\psi^\uparrow(j)}{H^{++}_{[x,L]} \psi^\uparrow(j)}
  \geq (A(\Delta) - \epsilon_\lambda(L-x+1,n-j)) \|\psi^\uparrow(j)\|^2\, .
$$
By the definition of $\epsilon_\lambda'(.)$ and $\epsilon_\lambda''(.)$, 
$$
\epsilon_\lambda(L-x+1,n-j)
 \leq \epsilon_\lambda'(n-j)
 \leq \epsilon_\lambda''(\ceil{n/2})\, ,
$$
since $n-j \geq \ceil{n/2}$.
So
\begin{equation}
\label{ROP:aneq}
\ip{\psi^\uparrow(j)}{H^{++}_{[x,L]} \psi^\uparrow(j)}
  \geq (A(\Delta) - \epsilon_\lambda''(\ceil{n/2})) \|\psi^\uparrow(j)\|^2\, .
\end{equation}
By Proposition \ref{kink-gap:Prop}, 
$$
\ip{\psi^\uparrow(j)}{H^{+-}_{[1,x]} \psi^\uparrow(j)} \\
  \geq \gamma \ip{\psi^\uparrow(j)}
  {\Big((\unity - \Proj(\psi^{+-}_{[1,x]}(j))\otimes \unity_{[x+1,L]}\Big)
  \psi^\uparrow(j)}\, .
$$
We can prove
\begin{equation}
\label{rop:little}
\ip{\psi^\uparrow(j)}{\Proj(\psi^{+-}_{[1,x]}(j)) \otimes \unity_{[x+1,L]}
  \psi^\uparrow(j))} 
  \leq \frac{q^{|J|}}{\pn \infty} \|\psi^\uparrow(j)\|^2\, .
\end{equation}
Indeed, since $\psi^\uparrow(j) \in \Hil_{[1,a-1],j} \otimes \Hil_{[a,x],0} 
\otimes \Hil_{[x+1,L],n-j}$
we have
\begin{eqnarray*}
&&\ip{\psi^\uparrow(j)}{\Proj(\psi^{+-}_{[1,x]}(j)) \otimes \unity_{[x+1,L]}
  \psi^\uparrow(j)} 
  \leq \|\psi^\uparrow(j)\|^2 \\
&&\quad\times
  \|\Proj(\psi^{+-}_{[1,x]}(j)) 
  \Proj(\Hil_{[1,a-1],j} \otimes \Hil_{[a,x],0}) \|^2\, ;
\end{eqnarray*}
so it suffices to check
$$
\|\Proj(\psi^{+-}_{[1,x]}(j)) 
  \Proj(\Hil_{[1,a-1],j} \otimes \Hil_{[a,x],0})\|^2 
  \leq \frac{q^{|J|}}{\pn \infty}\, .
$$
But, by a computation,
\begin{eqnarray*}
&&\|\Proj(\psi^{+-}_{[1,x]}(j))
  \Proj(\Hil_{[1,a-1],j} \otimes \Hil_{[a,x],0}) \| \\
&&\hspace{2cm} \displaystyle
  = \frac{\|\Proj(\Hil_{[1,a-1]}^{j} \otimes \Hil_{[a,x]}^{0}) 
  \psi^{+-}_{[1,x]}(j) \|^2}{\|\psi^{+-}_{[1,x]}(j)\|^2} \\
&&\hspace{2cm} \displaystyle
  = \frac{\|q^{j(x-a+1)} \psi^{+-}_{[1,a-1]}(j) \otimes 
  \psi^{+-}_{[a,x]}(0)\|^2}{\|\psi^{+-}_{[1,x]}(j)\|^2} \\
&&\hspace{2cm}
  = \qbinom{a-1}{j}{q^2} q^{2 j(\floor{|J|/2}+1)} \Big/
  \qbinom{x}{j}{q^2} \\
&&\hspace{2cm} \displaystyle
  \leq \frac{q^{|J|}}{\pn \infty}\, .
\end{eqnarray*}
The last calculation is deduced from equations \eq{App:+-coprod}
and \eq{App:-+coprod}, and note that it is necessary that $j \geq 1$.
From this we conclude
\begin{equation}
\ip{\psi^\uparrow(j)}{H^{+-}_{[1,x]} \psi^\uparrow(j)} 
  \geq \gamma(1 - \frac{q^{|J|}}{\pn \infty}) 
  \|\psi^\uparrow(j)\|^2\, .
\end{equation}
Combining this with \eq{ROP:aneq}, we have
\begin{equation}
\label{rop:3}
\begin{array}{l}
\ip{\psi^\uparrow(j)}{H^{++}_{[1,L]} \psi^\uparrow(j)} \vspace{2mm}\\
\hspace{1cm} \displaystyle
  \geq \left(A(\Delta) - \epsilon_\lambda''(\ceil{n/2})
  + \gamma \left(1 - \frac{q^{|J|}}{\pn \infty}\right)\right)
  \|\psi^\uparrow(j)\|^2
\end{array}
\end{equation}
as long as $1 \leq j \leq \floor{n/2}$.
A symmetric argument yields the same bound for the case
that $\ceil{n/2} \leq j \leq n-1$.

For $j = 0$, note that 
$\psi^\uparrow(0) = \ket{\uparrow \dots \uparrow}_{[1,x]} \otimes 
\psi'_{[x+1,L]}$, for some $\psi'_{[x+1,L]} \in \Hil_{[x+1,L],n}$.
Also, by \eq{useful1},
$$
H^{++}_{[1,L]} = H^{+-}_{[1,x+1]} + H^{++}_{[x+1,L]}
  \geq H^{++}_{[x+1,L]}\, .
$$
So
\begin{eqnarray*}
&\ip{\psi^\uparrow(0)}{H^{++}_{[1,L]} \psi^\uparrow(0)}
  &\geq \ip{\psi'_{[x+1,L]}}{H^{++}_{[x+1,L]} \psi'_{[x+1,L]}} \\
  &&\geq (A(\Delta) - \epsilon_\lambda(L-x,n)) \|\psi'_{[x+1,L]}\|^2 \\
  &&  + \lambda \ip{\psi'_{[x+1,L]}}{\Big(\unity 
  - \sum_{\tilde{x}=x+\floor{n/2}}^{L-\ceil{n/2}} 
  \Proj(\xi_{[x+1,L],n}(\tilde{x})\Big) \psi'_{[x+1,L]}}\, .
\end{eqnarray*}
We can replace $\|\psi'_{[x+1,L]}\|^2$ by $\|\psi^\uparrow(0)\|^2$.
Also, since 
$$
\psi'_{[x+1,L]} \in \Hil_{[x+1,b],0} \otimes \Hil_{[b+1,L],n}\, ,
$$
it is true that
$$
  \Proj(\xi_{[x+1,L]}(\tilde{x},n)) \psi'_{[x+1,L]}=0
$$
unless $\tilde{x} \geq b+\floor{n/2}$.
Furthermore, 
$\Proj(\Hil_{[1,x],0} \otimes \Hil_{[x+1,L],n})
\xi_{[x+1,L],n}(\tilde{x}) = 
\ket{\uparrow \dots \uparrow}_{[1,x]} \otimes \xi_{[x+1,L],n}(\tilde{x})$.
Therefore
\begin{eqnarray*}
&&\ip{\psi'_{[x+1,L]}}{\sum_{\tilde{x}=b+\floor{n/2}}^{L-\ceil{n/2}}
  \Proj(\xi_{[x+1,L],n}(\tilde{x})) \psi'_{[x+1,L]}}\\
&&\hspace{1cm}
  = \ip{\psi^\uparrow(0)}{\sum_{\tilde{x}=b+\floor{n/2}}^{L-\ceil{n/2}}
  \frac{\|\xi_{[x+1,L],n}(\tilde{x})\|^2}
  {\|\xi_{[1,L],n}(\tilde{x})\|^2}\Proj(\xi_{[1,L],n}(\tilde{x})) 
  \psi^\uparrow(0)}\, ,
\end{eqnarray*}
But it is very easy to see that 
$\|\xi_{[x+1,L],n}(\tilde{x})\|^2 \leq \|\xi_{[1,L],n}(\tilde{x})\|^2$.
So
\begin{equation}
\label{rop:4}
\begin{array}{l}
\displaystyle
\ip{\psi^\uparrow(0)}{H^{++}_{[1,L]} \psi^\uparrow(0)}
  \geq (A(\Delta) - \epsilon_\lambda'(\frac{3}{4}L,n))
  \|\psi^\uparrow(0)\|^2 \vspace{2mm}\\
\displaystyle
\hspace{1cm}
  + \lambda
  \ip{\psi^\uparrow(0)}
  {\Big(\unity - \sum_{\tilde{x} = b+\floor{n/2}}^{L-\ceil{n/2}} 
  \Proj(\xi_L(\tilde{x}))\Big) \psi^\uparrow(0)}\, .
\end{array}
\end{equation}
By an analogous argument
\begin{equation}
\label{rop:5}
\begin{array}{l}
\displaystyle
\ip{\psi^\uparrow(n)}{H^{++}_{[1,L]} \psi^\uparrow(n)}
  \geq (A(\Delta) - \epsilon_\lambda'(\frac{3}{4}L,n))
  \|\psi^\uparrow(n)\|^2  \vspace{2mm}\\
\displaystyle
\hspace{1cm}
  + \lambda
  \ip{\psi^\uparrow(n)}
  {\Big(\sum_{\tilde{x} = \floor{n/2}}^{a-1-\ceil{n/2}} 
  \Proj(\xi_{L,n}(\tilde{x}))\Big) \psi^\uparrow(n)}\, .
\end{array}
\end{equation}

Let us summarize the proof so far.
We began with a state $\psi \in \Hil_{L,n}$.
By Corollary \ref{IHS:Cor}, we found an interval $J$
such that $P_J \psi$ is a good approximation to $\psi$.
We decomposed $P_J \psi$ according to whether $\psi$ is in the
range of $P_J^\uparrow$ or $P_J^\downarrow$, and by the number of
downspins to the left of $J$.
We split the states $\psi^\sigma(j)$ into five classes
($\sigma = \downarrow$; $\sigma = \uparrow$, $j=0$;
$\sigma = \uparrow$, $1 \leq j \leq \floor{n/2}$;
$\sigma = \uparrow$, $\floor{n/2} \leq j \leq n-1$;
$\sigma = \uparrow$, $j=n$) and gave some spectral
gap estimates for each.
The only piece of the proof left is an induction argument, 
and one other thing: a proof that all of the spectral gap
estimates for each of the states $\psi^\sigma(j)$ can
be combined to a single spectral gap estimate for $P_J \psi$.
Specifically, while the $\psi^{\sigma}(j)$ are orthogonal
with respect to $\ip{*}{*}$ and $\ip{*}{H^{++}_{[1,L]}*}$,
it is not true that they are orthogonal with respect to
$\ip{*}{\Proj(\xi_{L,n}(\tilde{x}))*}$ for every $\tilde{x}$.
The trick is that they are nearly orthogonal with respect
to the projection for specific choices of $\tilde{x}$:
namely, if $\tilde{x} \in I_1 \cup I_2 \cup I_3$,
where $I_1 = [\floor{n/2},a-1-\ceil{n/2}]$, 
$I_2 = [a-1-\floor{n/2}+|J|,b+\floor{n/2}-|J|]$
and $I_3 = [b+\floor{n/2},L-\ceil{n/2}]$.
We will prove in Appendix B that,
in fact
\begin{eqnarray*}
&& \ip{P_J \psi}{\sum_{\tilde{x} \in I_1 \cup I_2 \cup I_3} 
  \Proj(\xi_{L,n}(x)) P_J \psi} \\
&& \qquad \geq - C_4(q) q^{|J|} \|P_J \psi\|^2
  + \sum_{j=0}^{n-|J|} \ip{\psi^\downarrow(j)}
  {\Proj(\xi_{L,n}(\tilde{x}_j)) \psi^\downarrow(j)} \\
&& \qquad + \sum_{\tilde{x}=b+\floor{n/2}}^{L-\ceil{n/2}}
  \ip{\psi^\uparrow(0)}{\Proj(\xi_{L,n}(\tilde{x})) \psi^\uparrow(0)} \\
&& \qquad + \sum_{\tilde{x}=\floor{n/2}}^{a-1-\ceil{n/2}}
  \ip{\psi^\uparrow(n)}{\Proj(\xi_{L,n}(\tilde{x})) \psi^\uparrow(n)}\, ,
\end{eqnarray*}
for some $C_4(q) < \infty$, as long as $n \geq N_4(q)$.

Equations \eq{rop:1}--\eq{rop:5} together with the result of
Appendix B imply
$$
\begin{array}{l}
\displaystyle
\ip{P_j \psi}{H^{++}_{[1,L]} P_J \psi}
  \geq (A(\Delta) - \eta) \|P_J \psi\|^2
\vspace{2mm}\\
\hspace{1cm} \displaystyle
  + \lambda \ip{P_J \psi}{\Big(\unity - \sum_{\tilde{x} \in I_1 + I_2 + I_3}
  \Proj(\xi_L(\tilde{x},n))\Big) P_J \psi}\, ,
\end{array}
$$
where
$$
\eta \leq (C_3(q) + C_4(q)) q^{|J|/2} 
  + \max\{0,\epsilon_\lambda''(\ceil{n/2}) - (\gamma-\lambda),
  \epsilon_\lambda'(\frac{3}{4}L,n)\}\, .
$$
Since each term $-\lambda \Proj(\xi_{L,n}(\tilde{x}))$, for
$\tilde{x} \in (I_1 \cup I_2 \cup I_3)'$
gives a negative contribution to the expectation, we can 
add those terms to the inequality:
\begin{equation}
\label{rop:7}
\begin{array}{rcl}
\displaystyle
\ip{P_j \psi}{H^{++}_{[1,L]} P_J \psi}
  &\geq&\displaystyle (A(\Delta) - \eta) \|P_J \psi\|^2
\vspace{1mm}\\
&&\displaystyle
  + \lambda \ip{P_J \psi}{\Big(\unity - \sum_{\tilde{x} = \floor{n/2}}
  ^{L - \ceil{n/2}}
  \Proj(\xi_L(\tilde{x},n))\Big) P_J \psi}\, ,
\end{array}
\end{equation}
Using \eq{Eval:result2}, and the fact that
$\|\unity - P\| \leq 1$, for any projection $P$, 
we have
$$
\Big\|\unity - \sum_{\tilde{x} = \floor{n/2}}^{L-\ceil{n/2}}
  \Proj(\xi_{L,n}(\tilde{x}))\Big\| 
  \leq 1 + \frac{2 q^n}{(1-q) \pn{\infty}}\, .
$$
This and \eq{ROP:corres1}, \eq{ROP:corres2}
and \eq{rop:7} imply 
\begin{eqnarray*}
\ip{\psi}{H^{++}_{[1,L]} \psi}
  &\geq& (A(\Delta) - \epsilon_\lambda(L,n)) \|\psi\|^2 \\
  && + \lambda \ip{\psi}{\Big(\unity - \sum_{\tilde{x} = \floor{n/2}}
  ^{L- \ceil{n/2}} \Proj(\xi_{L,n}(\tilde{x})) \Big) \psi}
\end{eqnarray*}
where, for some $C_5(q)$ and $C_6(q)$,
\begin{eqnarray*}
\epsilon_\lambda(L,n) 
  &\leq& \eta + A(\Delta) C_1(q) L^{-1/2} + C_2(q) L^{-1/4} \\
  && + 2 \lambda(1 + \frac{2q^n}{(1-q^n)f_q(\infty)}) 
  C_1(q)^{1/2} L^{-1/4} \\
  &\leq& C_5(q) q^{\frac{1}{2} \sqrt{L}} + C_6(q) L^{-1/4} \\ 
  && + \max\{0,\epsilon_\lambda''(\ceil{n/2}) - (\gamma-\lambda),
  \epsilon_\lambda'(\frac{3}{4}L,n)\}\, .
\end{eqnarray*}
We have not stated the exact dependence of $C_5(q)$ and $C_6(q)$ on
$q$, though it can be deduced from our previous calculations.
The important fact is that there exists $N_5(q)$, such that
if $n \geq N_5(q)$, then the above holds with $C_5(q)$ and $C_6(q)$
both finite, positive numbers.
From this, it follows
$$
\epsilon_\lambda'(\frac{4}{3}L,n)
  \leq C_5(q) q^{\frac{1}{2} \sqrt{L}}
  + C_6(q) L^{-1/4} 
  + \max\{0,\epsilon_\lambda''(\ceil{n/2}) + \lambda - \gamma,
  \epsilon_\lambda'(L,n)\}\, ,
$$
and
\begin{eqnarray*}
\epsilon_\lambda'((\frac{4}{3})^k n,n)
  &\leq& C_5(q) q^{\frac{1}{2} \sqrt{n}} 
  \sum_{r=1}^{k-1} q^{[(4/3)^{r/2}-1]\sqrt{n}}
  + C_6(q) n^{-1/4} \sum_{r=1}^{k-1} (\frac{3}{4})^{r/4}\\
  &&\quad+ \max\{0,\epsilon_\lambda''(\ceil{n/2}) + \lambda - \gamma,
  \epsilon_\lambda'(n,n)\}\, .
\end{eqnarray*}
Note $\epsilon_\lambda'(n,n)=0$, because $\Hil_{n,n}$ is one-dimensional,
and the single vector $\xi_{n,n}(\floor{n/2}) 
= \ket{\downarrow \dots \downarrow}$ satisfies
$H^{++}_{[1,L]} \xi_n(\floor{n/2},n) 
= A(\Delta) \xi_n(\floor{n/2},n)$.
Therefore,
\begin{eqnarray*}
\epsilon_\lambda'(\infty,n)
  &\leq& C_5 q^{\frac{1}{2} \sqrt{n}} \sum_{k=1}^{\infty} 
  q^{[(4/3)^{k/2}-1]\sqrt{n}}
  + C_6 n^{-1/4} \sum_{k=1}^{\infty} (\frac{3}{4})^{k/4}\\
  &&\quad + \max\{0,\epsilon_\lambda''(\ceil{n/2}) + \lambda - \gamma\}\, .
\end{eqnarray*}
Taking the $\limsup$ as $n \to \infty$, we find
$$
\epsilon_\lambda''(\infty)
  \leq \max\{0,\epsilon_\lambda''(\infty) + \lambda - \gamma\}\, .
$$
For $\lambda < \gamma$ this implies $\epsilon_\lambda''(\infty)$ either
equals zero or $+\infty$.
But, by Proposition \ref{APBound:Prop},
$\epsilon''_\lambda(\infty)<A(\Delta)$.
So $\epsilon''_\lambda(\infty) = 0$, as desired, for every $\lambda<\gamma$.

By the Cantor diagonal argument, there is a sequence
$\epsilon_n$ satisfying \eq{ROP:piece1},
constructed from the $\epsilon_\lambda(n)$, with $\lambda \to \gamma$
and $n \to \infty$.
So Theorem \ref{main:theorem}(a) is proved.
Theorem \ref{main:theorem2} is a reformulation of the same result, 
so it needs no proof.

%%%%%%%%%%%%%%%%%%%%%%%%%%%%%%%%%%%%%%%%%%%%%%%%%%%%%%%%%%%%%%%%%%%%%%%%%%%%%%%%
%%%%%%%%%%%%%%%%%%%%%%%%%%%%%%%%%%%%%%%%%%%%%%%%%%%%%%%%%%%%%%%%%%%%%%%%%%%%%%%%
%%%%%%%%%%%%%%% Periodic -- Infinite        %%%%%%%%%%%%%%%%%%%%%%%%%%%%%%%%%%%%
%%%%%%%%%%%%%%%%%%%%%%%%%%%%%%%%%%%%%%%%%%%%%%%%%%%%%%%%%%%%%%%%%%%%%%%%%%%%%%%%
%%%%%%%%%%%%%%%%%%%%%%%%%%%%%%%%%%%%%%%%%%%%%%%%%%%%%%%%%%%%%%%%%%%%%%%%%%%%%%%%
\section{Results for the Ring and the Infinite Chain}
\label{Sec:Periodic-Infinite}

\subsection{The Spin Ring}

The spin ring (periodic spin chain) has state space
$\Hil_{L}$ and is defined by the Hamiltonian
$
\HXXZ_{\Ir/L} = \sum_{x=1}^{L-1} \HXXZ_{x,x+1} + \HXXZ_{L,1}\, .
$
We define a periodic droplet with $n$ down spins
$$
\xi_{\Ir/L,n}(0) = \xi_{L,n}(\floor{L/2})
  = \psi^{+-}_{[1,\floor{L/2}]}(\floor{n/2})
  \otimes \psi^{-+}_{[\floor{L/2}+1,L]}(\ceil{n/2})\, .
$$
There are $L-1$ additional droplet states
$$
\xi_{\Ir/L,n}(x) = T^x \xi_{\Ir/L,n}(0)
\qquad (x = 1,\dots,L-1)
$$
where $T$ is the unitary operator on $\Hil_L$
such that
$$ 
T(v_1 \otimes v_2 \otimes \dots \otimes v_L)
  = v_L \otimes v_1 \otimes \dots \otimes v_{L-1}\, .
$$
Let $\mathcal{K}_{\Ir/L,n}$ be the span of
$\xi_{\Ir/L,n}(x),\dots,\xi_{\Ir/L}(L-1,n)$.
Let 
$$
 \lambda_{\Ir/L,n}(1) \leq \lambda_{\Ir/L,n}(2) \leq 
 \dots \lambda_{\Ir/L,n}(\sbinom{L}{n})
$$
be the ordered eigenvalues of $\HXXZ_{\Ir/L}$ acting on the invariant
subspace $\Hil_{L,n}$, and let $\Hil_{\Ir/L,n}^k$ be the span of the
first $k$ eigenvectors.

\begin{theorem}
\label{Thm:Periodic}
For $1 \leq n \leq L-1$
$$
  \lambda_{\Ir/L,n}(1),\dots,\lambda_{\Ir/L,n}(L)
  \in [2 A(\Delta) - O(q^{n}+q^{L-n}),2 A(\Delta) + O(q^n + q^{L-n})]\, .
$$
Also,
$$
\liminf_{n,L \atop \min(n,L-n) \to \infty}
  \lambda(L,n,L+1) \geq 2 A(\Delta) + \gamma\, .
$$
Finally,
$$
\|\Proj(\mathcal{K}_{\Ir/L,n}) - \Proj(\Hil_{\Ir/L,n}^L)\|
  = O(q^n + q^{L-n})\, .
$$
\end{theorem}

\begin{proof}
We first prove that 
\begin{equation}
\label{pc:1}
\|(\HXXZ_{\Ir/L} - 2 A(\Delta)) \Proj(\calK_{\Ir/L,n})\| 
= O(q^{n} + q^{L-n})\, .
\end{equation}
It is easy to see that, just as for the droplets on an interval,
\begin{eqnarray}
\label{Per:1}
\frac{|\ip{\xi_{\Ir/L,n}(x)}{\xi_{\Ir/L,n}(y)}|}
  {\|\xi_{\Ir/L,n}(x)\| \cdot \|\xi_{\Ir/L,n}(y)\|}
  &\leq& \frac{q^{n\cdot d(x,y)}}{\pn \infty} 
  \quad \textrm{for all} \quad x,y\, ;\\
\label{Per:2}
\frac{|\ip{\xi_{\Ir/L,n}(x)}{\HXXZ_{\Ir/L} \xi_{\Ir/L,n}(y)}|}
  {\|\xi_{\Ir/L,n}(x)\| \cdot \|\xi_{\Ir/L,n}(y)\|}
  &\leq& \frac{q^{n\cdot d(x,y)}}{\pn \infty} 
  \quad \textrm{if} \quad x\neq y\, ;\\
\label{Per:3}
\frac{|\ip{\xi_{\Ir/L,n}(x)}{(\HXXZ_{\Ir/L})^2 \xi_{\Ir/L,n}(y)}|}
  {\|\xi_{\Ir/L,n}(x)\| \cdot \|\xi_{\Ir/L,n}(y)\|}
  &\leq& \frac{q^{n\cdot d(x,y)}}{\pn \infty} 
  \quad \textrm{if} \quad d(x,y) \geq 2\, ;
\end{eqnarray}
where $d(x,y) = \min(|x-y|,|x+y-L|)$.
In fact, using the same tools as in Appendix A, 
we can calculate exactly, for $0\leq x\leq \floor{L/2}$,  
$$
\rho(\xi_{\Ir/L,n}(0),T^x)
  = q^{n x} \sum_k \frac{\sqbinom{\floor{L/2}-x}{\floor{n/2}-k}{q^2}
  \sqbinom{\ceil{L/2}-x}{\ceil{n/2}-k}{q^2}}
  {\sqbinom{\floor{L/2}}{\floor{n/2}}{q^2}
  \sqbinom{\ceil{L/2}}{\ceil{n/2}}{q^2}}
  \binom{x}{k}^2 q^{k(L+2k)}\, .
$$
It is verifiable that this satisfies the bounds above.
The other expectations $\rho(\xi_{\Ir/L,n}(0),\HXXZ_{\Ir/L,n} T^x)$
and $\rho(\xi_{\Ir/L,n}(0),(\HXXZ_{\Ir/L,n})^2 T^x)$
are similar.
Applying Lemma \ref{Lem:OrthStates}, proves \eq{pc:1}.

Now we prove that, considering $\HXXZ_{\Ir/L}$ acting on the invariant subspace
$\Hil_{\Ir/L,n}$,
\begin{equation}
\label{pc:2}
\HXXZ_{\Ir/L} \geq (2 A(\Delta) - \epsilon_n - \epsilon_{L-n}) \unity
  + \gamma (\unity - \Proj(\calK_{L,n}))\, ,
\end{equation}
where $\lim_{n \to \infty} \epsilon_n = 0$.
To do this, we use Corollary \ref{IHS:Cor}.
There exists an $L_0(q)$ and $C_0(q)$
such that, if  $L > L_0(q)$ 
then for any 
$\psi \in \Hil_{L,n}$ with $\rho(\psi,\HXXZ_{\Ir/L}) \leq 2 A(\Delta) + \gamma$,
Corollary \ref{IHS:Cor} guarantess the existence of a ``subinterval'' 
$J \subset \Ir/L$ satisfying
$|J| = 2 \floor{L^{1/2}}$, $\|P_J \psi - \psi\| \leq C_0(q) L^{-1/2}$,
and
$$
\ip{\psi}{\HXXZ_{\Ir/L} \psi}
  \geq \ip{P_J \psi}{\HXXZ_{\Ir/L} P_J \psi}
  - C_0(q) L^{-1/4} \|\psi\|^2\, .
$$
We can take
$L_0(q) = (7-6q+3q^2)^2/(1-q)^4$ 
and
$C_0(q) =(5 + 18 q + 5 q^2) L_0(q)^{1/4}/(2+2q^2)$.
By ``subinterval'', we mean that there exists an interval $J' \subset \Ir$,
such that $J \equiv J' ({\rm mod} L)$.
Without loss of generality,
we assume $J = [1,\dots,\floor{L^{1/2}}] \cup [L+1-\floor{L^{1/2}},L]$.
Next,
$$
\ip{P_J \psi}{\HXXZ_{\Ir/L} P_J \psi}
  = \ip{P_J^\uparrow \psi}{\HXXZ_{\Ir/L} P_J^\uparrow \psi}
  + \ip{P_J^\downarrow \psi}{\HXXZ_{\Ir/L} P_J^\downarrow \psi}\, 
$$
and $\|P_J \psi\|^2 = \|P_J^\uparrow \psi\|^2 + \|P_J^\downarrow \psi\|^2$.

We estimate $\ip{P_J^\uparrow \psi}{\HXXZ_{\Ir/L} P_J^\uparrow \psi}$, first.
Of course, $\HXXZ_{\Ir/L} = H^{--}_{L,1} + H^{++}_{[1,L]}$,
and since $H^{--} \ket{\uparrow \uparrow} = A(\Delta) \ket{\uparrow \uparrow}$,
we see that
$
\HXXZ_{\Ir/L} P_J^\uparrow \psi 
  = (A(\Delta) + H^{++}_{[1,L]}) P_J^\uparrow \psi
$.
Then using Theorem \ref{main:theorem}(b),
\begin{eqnarray*}
&&\ip{P_J^\uparrow \psi}{\HXXZ_{\Ir/L} P_J^\uparrow \psi}
  \geq (A(\Delta) - \epsilon(n)) \|P_J^\uparrow \psi\|^2 \\
&&\qquad  \qquad + \gamma
  \ip{P_J^\uparrow \psi}
  {(\unity - \Proj(\calK_{L,n}))P_J^\uparrow \psi}\, ,
\end{eqnarray*}
where $\lim_{n \to \infty} \epsilon(n) = 0$.
But 
\begin{eqnarray*}
P_J^\uparrow \Proj(\calK_{L,n}) P_J^\uparrow
  &=& P_J^\uparrow \sum_{x=\floor{n/2}}^{L-\ceil{n/2}} \Proj(\xi_{L,n}(x))
  P_J^\uparrow + O(q^n) \\
  &=& P_J^\uparrow \sum_{x=\floor{L^{1/2}}+\floor{n/2}}
  ^{L+1-\floor{L^{1/2}}-\ceil{n/2}} \Proj(\xi_{L,n}(x)) P_J^\uparrow
  + O(q^n) \\
  &\leq& P_J^\uparrow \sum_{x=0}^{L-1} \Proj(\xi_{\Ir/L,n}(x)) P_J^\uparrow
  - O(q^n) \\
  &=& P_J^\uparrow \Proj(\calK_{\Ir/L,n}) P_J^\uparrow - O(q^n)\, ,
\end{eqnarray*}
where by $A = B + O(q^n)$, we mean $\|A - B\| = O(q^n)$,
and by $A \geq B - O(q^n)$, we mean $B-A \leq O(q^n) \unity$.
We omit the calculations here.
So
\begin{eqnarray*}
&& \ip{P_J^\uparrow \psi}{\HXXZ_{\Ir/l} P_J^\uparrow \psi}
  \geq (2A(\Delta) - \epsilon(n) - O(q^n)) \|P_J^\uparrow \psi\|^2 \\
&&\qquad   + \gamma \ip{P_J^\uparrow \psi}{(\unity - \Proj(\calK_{\Ir/L,n})) 
  P_J^\uparrow \psi}\, .
\end{eqnarray*}
Symmetrically,
\begin{eqnarray*}
&& \ip{P_J^\downarrow \psi}{\HXXZ_{\Ir/l} P_J^\downarrow \psi}
  \geq (2A(\Delta) - \epsilon(L-n) - O(q^{L-n})) \|P_J^\downarrow \psi\|^2 \\
&&\qquad   + \gamma \ip{P_J^\downarrow \psi}{(\unity - 
  \Proj(F \calK_{\Ir/L,L-n})) P_J^\downarrow \psi}\, ,
\end{eqnarray*}
where $F : \Hil_{L,L-n} \to \Hil_{L,n}$ denotes the spin-flip.
But $\calK_{\Ir/L,n} = F \calK_{\Ir/L,L-n}$.
Also,
$\|P_J^\downarrow \Proj(\calK_{\Ir/L,n}) P_J^\uparrow\| = O(q^n + q^{L-n})$.
So, for any $\psi \in \Hil_{L,n}$,
\begin{equation}
\label{pc:3}
\begin{array}{l}
\displaystyle
\ip{\psi}{\HXXZ_{\Ir/l} \psi} \vspace{2mm}\\
\qquad \displaystyle
  \geq (2A(\Delta) - [\epsilon_n + \epsilon_{L-n} 
  + O(q^n+q^{L-n}) + O(L^{-1/4})]) 
  \|\psi\|^2 \vspace{2mm}\\
\qquad \qquad \displaystyle  
+ \gamma \ip{\psi}{(\unity - \Proj(\calK_{\Ir/L,n})) \psi}\, .
\end{array}
\end{equation}
Equations \eq{pc:1} and \eq{pc:3} together imply the corollary.
\end{proof}

\subsection{The Infinite Spin Chain}

Let $\ket{\Omega} = \ket{\dots \uparrow \uparrow \uparrow \dots}_{\Ir}$ be a 
vacuum state, and define 
$$
\Hil_{\Ir,n} = {\rm cl}( 
\Span\{S_{x_1}^- S_{x_2}^- \dots S_{x_n}^- \ket{\Omega}
  : x_1 < x_2 < \dots < x_n\})\, ,
$$
where ${\rm cl}(.)$ is the $l^2$-closure.
This is a separable Hilbert space, and 
$$
\HXXZ_\Ir = \sum_{x=-\infty}^\infty \HXXZ_{x,x+1}
$$
is a densely defined, self-adjoint operator.
This Hamiltonian defines the infinite spin chain.
We check that the series does converge.
In fact
$$
0 \leq \HXXZ_{x,x+1} \leq \frac{1}{2}(1+\Delta^{-1})(\hat{N}_x + \hat{N}_{x+1})
$$
where $\hat{N}_x = (\frac{1}{2} - S_x^3)$ counts the number of down spins at 
$x$. But $\sum_{x=-\infty}^\infty \hat{N}_x \equiv n$ on $\Hil_{\Ir,n}$.
So the series does converge, and
$\HXXZ_\Ir \leq n (1+\Delta^{-1})$.
We define the droplet states
$$
\xi_{\Ir,n}(x) = \psi^{+-}_{(-\infty,x]}(\floor{n/2})
  \otimes \psi^{-+}_{[x,\infty)}(\ceil{n/2});
$$
and let $\calK_{\Ir,n}$ be the $l^2$ closure of 
$\Span\{\xi_{\Ir,n}(x) : x \in \Ir\})$.

\begin{theorem}
\label{Thm:Infinite}
The following bounds exist for the infinite spin chain
$$
\|(\HXXZ_\Ir - 2 A(\Delta)) \Proj(\calK_{\Ir,n})\|
  = O(q^n)\, ,
$$
and, considering $\HXXZ_\Ir$ as an operator on $\Hil_{\Ir,n}$,
$$
\HXXZ_\Ir \geq (2 A(\Delta) - \epsilon_n) \unity
  + \gamma (\unity - \Proj(\calK_{\Ir,n}))\, ,
$$
where $\epsilon_n$ is a sequence with $\lim_{n \to \infty} \epsilon_n = 0$.
\end{theorem}

\begin{proof}
The proof that 
\begin{equation}
\label{ic:1}
\|(\HXXZ_{\Ir} - 2 A(\Delta)) \Proj(\calK_{\Ir,n})\| = O(q^n)\,
\end{equation}
is essentially the same as in Section \ref{Sect:Eval}.
One fact we should check is that for each $\xi_{\Ir,n}(x)$,
$\|(\HXXZ_\Ir - 2 A(\Delta)) \xi_{\Ir,n}(x)\|^2 = O(q^n)$.
We observe that
\begin{eqnarray*}
\HXXZ_{[-L,L]} \xi_{\Ir,n}(0)
  &=& (H^{+-}_{[-L,0]} + H^{-+}_{[1,L]} + H^{++}_{0,1} 
  + A(\Delta) (S_{-L}^3 + S_L^3))
  \xi_{\Ir,n}(0) \\
  &=& (H^{++}_{0,1} + A(\Delta) (S_{-L}^3 + S_L^3)) \xi_{\Ir,n}(0)\, .
\end{eqnarray*}
But as before, 
$$
\|(H^{++}_{0,1} - A(\Delta)) \xi_{\Ir,n}(0)\|^2 
  \leq O(q^n) \|\xi_{\Ir,n}(0)\|^2\, .
$$
An obvious fact is
$$
\|(S_{-L}^3 + S_L^3 - 1) \xi(0,n)\|^2 \leq O(q^{L-n}) \|\xi(0,n)\|^2\, .
$$
Taking $L \to \infty$, yields the desired result.
We have the usual orthogonality estimates 
\begin{eqnarray*}
\frac{|\ip{\xi_{\Ir,n}(x)}{\xi_{\Ir,n}(y)}|}
{\|\xi_{\Ir,n}(x)\| \cdot \|\xi_{\Ir,n}(y)\|}
  &\leq& \frac{q^{n|x-y|}}{\pn{\infty}} ,\\
\frac{|\ip{\xi_{\Ir,n}(x)}{\HXXZ_\Ir \xi_{\Ir,n}(y)}\|}
{\|\xi_{\Ir,n}(x)\| \cdot \|\xi_{\Ir,n}(y)\|}
  &\leq& \frac{q^{n|x-y|}}{\pn{\infty}}\quad
\textrm{for } x\neq y\, ,\\
\frac{|\ip{\xi_{\Ir,n}(x)}{(\HXXZ_\Ir)^2 \xi_{\Ir,n}(y)}|}
{\|\xi_{\Ir,n}(x)\| \cdot \|\xi_{\Ir,n}(y)\|}
  &\leq& \frac{q^{n|x-y|}}{\pn{\infty}}\quad
\textrm{for } |x-y| \geq 2\, .\\
\end{eqnarray*}
In fact, the estimate of $\ip{\xi_{\Ir,n}(x)}{\xi_{Ir,n}(y)}$
follows by \eq{App:result1}, taking the limit that $L \to \infty$,
and the other estimates are consequences.
Applying Lemma \ref{Lem:OrthStates} proves \eq{ic:1}.

For the second part, suppose $\psi \in \Hil_{\Ir,n}$.
Then 
$$
\rho(\psi,\HXXZ_\Ir) = \lim_{L \to \infty} \rho(\psi,\HXXZ_{[-L,L]})\, .
$$
Furthermore 
$\HXXZ_{[-L,L]} = H^{++}_{[-L,L]} + A(\Delta) (S_{-L}^3 + S_L^3)$,
and 
$$
\lim_{L \to \infty} \ip{\psi}{(S_{-L}^3 + S_L^3) \psi} = \|\psi\|^2
$$ 
by virtue of the fact that $n$, the total number of down spins 
in the state $\psi$,
is finite.
Essentially the same fact is restated as
$ \lim_{L \to \infty} \psi_L = \psi$, where
$$
\psi_L = 
  \Proj(\Hil_{(-\infty,-L-1],0} \otimes \Hil_{[-L,L],n} 
  \otimes \Hil_{[L+1,\infty),0}) \psi\, .
$$
Let us define 
$$
\Xi_{L,n} = 
  \Proj(\Hil_{(-\infty,-L-1],0} \otimes \mathcal{K}_{[-L,L],n} 
  \otimes \Hil_{[L+1,\infty),0}) \psi\, ,
$$
where $\mathcal{K}_{[-L,L],n}$ is the droplet state subspace
for the finite chain.
By Theorem \ref{main:theorem}(b),
\begin{eqnarray*}
&& \ip{\psi_L}{H^{++}_{[-L,L]} \psi_L} 
  \geq (2 A(\Delta) - \epsilon(n)) \|\psi_L\|^2 \\
&&\qquad \qquad  + \gamma \ip{\psi_L}
  {(\unity - \Xi_{L,n})\psi_L}\, .
\end{eqnarray*}
Since $\psi_L \to \psi$ in the norm-topology, as $L \to \infty$, 
all we need to check is that  
$\Xi_{L,n}$ converges weakly to $\Proj(\calK_{\Ir,n})$.

It helps to break up $\Xi_{L,n}$ into two pieces, 
\begin{eqnarray*}
&&
\Xi'_{L,n} = 
  \Proj(\Span\{ \ket{\dots \uparrow}_{(-\infty,-L-1]} 
  \otimes \xi_{[-L,L],n}(x)
  \otimes \ket{\uparrow \dots}_{[L+1,\infty)} :\\
&& \hspace{4cm}
  - \floor{L/2} + \floor{n/2} \leq x \leq \ceil{L/2} - \ceil{n/2} \})\, ,
\end{eqnarray*}
and $\Xi''_{L,n} = \Xi_{L,n} - \Xi'_{L,n}$.
Define
$$
\phi_{L,n}(x) = \ket{\dots \uparrow}_{(-\infty,-L-1]} 
  \otimes \xi_{[-L,L],n}(x)
  \otimes \ket{\uparrow \dots}_{[L+1,\infty)}\, .
$$
Note that for any sequence $x_L$ such that
$x_L \in [-\floor{L/2}+\floor{n/2},\ceil{L/2}-\ceil{n/2}]$,
we have
$$
\lim_{L \to \infty} \rho(\phi_{L,n}(x_L),\calK_{\Ir,n}) = 1\, .
$$
The reason is that $\|\phi_{L,n}(x) - \xi_\Ir(x,n)\| = O(q^{L/2})$ 
because the the left and right interfaces of the droplet in
$\phi_{L,n}(x)$ are a distance at least $L/2$ from the left and right endpoints
of the interval $[-L,L]$, and 
the probability of finding an overturned spin decays
$q$-exponentially with the distance from the inteface.
For the same reason, for any fixed $x \in \Ir$,
$\lim_{L,\to \infty} \rho(\xi_\Ir(x,n),\Xi'_{L,n}) = 1$.
These two facts imply that $\Xi'_{L,n}$ converges weakly to 
$\Proj(\calK_{\Ir,n})$.
Now $\Xi''_{L,n}$ converges weakly to zero, 
because every state in $\Xi''_{L,n}$ has over
half its downspins concentrated in the annulus
$[-L,L] \setminus [-\floor{L/2}+\floor{n/2},\ceil{L/2}-\ceil{n/2}]$,
and the inner radius tend to infinity.
This means that $\textrm{w}-\lim_{L \to \infty} \Xi_{L,n} = 
\Proj(\calK_{\Ir,n})$,
as claimed.

Thus, taking the appropriate limits,
\begin{eqnarray*}
&&\ip{\psi}{\HXXZ_\Ir \psi}
  \geq (2 A(\Delta) - \epsilon(n)) \|\psi\|^2 \\
&& \qquad  
  + \gamma \ip{\psi}{(\unity - \Proj(\calK_{\Ir,n})) \psi}\, ,
\end{eqnarray*}
which finishes the proof of the theorem.
\end{proof}

%%%%%%%%%%%%%%%%%%%%%%%%%%%%%%%%%%%%%%%%%%%%%%%%%%%%%%%%%%%%%%%%%%%%%%%%%%%%%%%
%%%%%%%%%%%%%%%%%%%%%%%%%%%%%%%%%%%%%%%%%%%%%%%%%%%%%%%%%%%%%%%%%%%%%%%%%%%%%%%
%%%%%%%%%%%%%%%%%%%%%%%%%%%%%%%%%%%%%%%%%%%%%%%%%%%%%%%%%%%%%%%%%%%%%%%%%%%%%%%

\section*{Appendix A}
\label{App}
\setcounter{equation}{0}
\renewcommand{\theequation}{A.\arabic{equation}}
In this section we carry out several calculations,
whose results are needed in the main body of the paper,
but whose proofs are not very enlightening for understanding
the main arguments.
The definitions of the kink states, $\psi^{+-}_{[a,b]}(n)$, and the
antikink states, $\psi^{-+}_{[a,b]}(n)$, are given in 
\eq{Intro:+-} and \eq{Intro:-+}.
One nice feature of these states is that they are governed by a quantum
Clebsh-Gordan formula, due to the $SU_q(2)$ symmetry of
$H^{\alpha \beta}_{[a,b]}$, $\alpha \beta = +-,-+$. 
By this we mean the following: Suppose $a\leq x\leq b$.
Then, 
\begin{eqnarray}
\label{App:+-coprod}
\psi^{+-}_{[a,b]}(n) &=& \sum_k \psi^{+-}_{[a,x]}(k) \otimes 
  \psi^{+-}_{[x+1,b]}(n-k) q^{(b-x)k}\, ,\\
\label{App:-+coprod}
\psi^{-+}_{[a,b]}(n) &=& \sum_k \psi^{-+}_{[a,x]}(k) \otimes 
  \psi^{-+}_{[x+1,b]}(n-k) q^{(x+1-a)(n-k)} \, .
\end{eqnarray}
We let the sum in $k$ run over all integers $k$, with the understanding that
$\psi^{+-}_{[a,b]}(n) = \psi^{-+}_{[a,b]}(n)$ if $n<0$ or $n>b-a+1$.
One need not refer to the quantum group to understand this decomposition,
it is enough just to check the definitions.
We can also see from the definitions that 
\begin{eqnarray}
\label{App:kink-norm}
\ip{\psi^{\alpha \beta}_{[a,b]}(m)}{\psi^{\alpha \beta}_{[a,b]}(n)}
  &=& \delta_{m,n} \qbinom{b-a+1}{n}{q^2} q^{n(n+1)} \\
\label{App:mixed-ip}
  \ip{\psi^{\alpha \beta}_{[a,b]}(m)}{\psi^{\beta \alpha}_{[a,b]}(n)}
  &=& \delta_{m,n} \binom{b-a+1}{n} q^{b-a+2}\, ,
\end{eqnarray}
for $\alpha \beta = +-,-+$.

The combinatorial prefactor in \eq{App:kink-norm}
is a $q$-binomial coefficient (in this case a $q^2$-binomial coefficient), 
also known as a Gauss polynomial.
The most important feature, for us, is the $q$-binomial formula
$$
\prod_{k=1}^L (1 + q^{2k} x) = \sum_{n=0}^L \qbinom{L}{n}{q^2} q^{n(n+1)}x^n \, .
$$
At this point let us introduce another useful combinatorial quantity,
$\pn n$, defined for $n=0,1,2,\dots,\infty$:
$$
\pn n = \prod_{k=1}^n (1 - q^{2k})\, .
$$
For a fixed $q \in [0,1)$, the sequence $\pn n$ is clearly montone decreasing,
and $\pn \infty > 0$. 
We note that 
$$
\qbinom{n}{k}{q^2} = \frac{\pn{n}}{\pn{k} \pn{n-k}}
$$
which means that for $0 \leq k \leq n$,
$$
1 \leq \qbinom{n}{k}{q^2} \leq \frac{1}{\pn \infty}\, .
$$
The first result we wish to prove is that
\begin{equation}
\label{App:prelim}
\begin{array}{l}
\displaystyle
\ip{\psi^{+-}_{[1,x]}(m) \otimes \psi^{-+}_{[x+1,x+y+r]}(n+k)}
{\psi^{+-}_{[1,x+r]}(m+k) \otimes \psi^{-+}_{[x+r+1,x+y+r]}(n)}
\vspace{2mm}  \\
\hspace{3cm} 
= \binom{r}{k} \qbinom{x}{m}{q^2} \qbinom{y}{n}{q^2}
  q^{m(m+k+1) + n(n+k+1) + k(r+1)}\, .
\end{array}
\end{equation}
This is very simple.
From \eq{App:+-coprod} and \eq{App:-+coprod},
\begin{eqnarray*}
&& \ip{\psi^{+-}_{[1,x]}(m) \otimes \psi^{-+}_{[x+1,x+y+r]}(n+k)}
  {\psi^{+-}_{[1,x+r]}(m+k) \otimes \psi^{-+}_{[x+r+1,x+y+r]}(n)} \\
&& \quad =
  \sum_{j,l} \bra{q^{r(n+k-j)} \psi^{+-}_{[1,x]}(m) \otimes 
  \psi^{-+}_{[x+1,x+r]}(j)
  \otimes \psi^{+-}_{[x+r+1,x+y+r]}(n+k-j)} \\
&& \qquad \qquad
  \ket{q^{r(m+k-l)} \psi^{+-}_{[1,x]}(l) \otimes \psi^{+-}_{[x+1,x+r]}(m+k-l)
  \otimes \psi^{-+}_{[x+r+1,x+y+r]}(n)} \\
&& \quad = \sum_{j,l} q^{r(m+n+2k-l-j)} 
  \ip{\psi^{+-}_{[1,x]}(m)}{\psi^{+-}_{[1,x]}(l)} \\
&& \qquad \qquad
  \times \ip{\psi^{-+}_{[x+1,x+r]}(j)}{\psi^{+-}_{[x+1,x+r]}(m+k-l)} \\
&& \qquad \qquad
  \times \ip{\psi^{-+}_{[x+r+1,x+y+r]}(n+k-j)}{\psi^{-+}_{[x+r+1,x+y+r]}(n)}\, .
\end{eqnarray*}
Consulting \eq{App:kink-norm} and \eq{App:mixed-ip}, we see that the only 
choice
of $j$ and $l$ for which none of the inner-products vanishes is $j=l=k$.
Plugging in these values for $j$ and $l$ and using the formulae for the 
inner-products yields \eq{App:prelim}.
We can use \eq{App:kink-norm} to normalize the inner-product in the following 
way,
\begin{equation}
\label{App:prelim2}
\begin{array}{l}
\displaystyle
\frac{ \ip{\psi^{+-}_{[1,x]}(m) \otimes \psi^{-+}_{[x+1,x+y+r]}(n+k)}
{\psi^{+-}_{[1,x+r]}(m+k) \otimes \psi^{-+}_{[x+r+1,x+y+r]}(n)} }
{\|\psi^{+-}_{[1,x]}(m) \otimes \psi^{-+}_{[x+1,x+y+r]}(n+k)\| \cdot
\|\psi^{+-}_{[1,x+r]}(m+k) \otimes \psi^{-+}_{[x+r+1,x+y+r]}(n)\|}
\vspace{2mm}  \\
\hspace{1cm} 
= \binom{r}{k} \sqrt{\qbinom{x}{m}{q^2} \qbinom{y}{n}{q^2}
 \Big/ \qbinom{x+r}{m+k}{q^2} \qbinom{y+r}{n+k}{q^2}}\,
  q^{(m+n+k)(r-k)}\, .
\end{array}
\end{equation}

We wish to specialize this formula in two ways.
First, by setting $k=r$ we have
\begin{equation}
\label{App:special1}
\begin{array}{l}
\displaystyle
\frac{ \ip{\psi^{+-}_{[1,x]}(m) \otimes \psi^{-+}_{[x+1,x+y+r]}(n+r)}
{\psi^{+-}_{[1,x+r]}(m+r) \otimes \psi^{-+}_{[x+r+1,x+y+r]}(n)} }
{\|\psi^{+-}_{[1,x]}(m) \otimes \psi^{-+}_{[x+1,x+y+r]}(n+r)\| \cdot
\|\psi^{+-}_{[1,x+r]}(m+r) \otimes \psi^{-+}_{[x+r+1,x+y+r]}(n)\|}
\vspace{2mm}  \\
\hspace{1cm} 
= \sqrt{\qbinom{x}{m}{q^2} \qbinom{y}{n}{q^2}
 \Big/ \qbinom{x+r}{m+r}{q^2} \qbinom{y+r}{n+r}{q^2}}\, .
\end{array}
\end{equation}
Second, by setting $k=0$, we have
\begin{equation}
\label{App:special2}
\begin{array}{l}
\displaystyle
\frac{ \ip{\psi^{+-}_{[1,x]}(m) \otimes \psi^{-+}_{[x+1,x+y+r]}(n)}
{\psi^{+-}_{[1,x+r]}(m) \otimes \psi^{-+}_{[x+r+1,x+y+r]}(n)} }
{\|\psi^{+-}_{[1,x]}(m) \otimes \psi^{-+}_{[x+1,x+y+r]}(n)\| \cdot
\|\psi^{+-}_{[1,x+r]}(m) \otimes \psi^{-+}_{[x+r+1,x+y+r]}(n)\|}
\vspace{2mm}  \\
\hspace{1cm} 
= \sqrt{\qbinom{x}{m}{q^2} \qbinom{y}{n}{q^2}
 \Big/ \qbinom{x+r}{m}{q^2} \qbinom{y+r}{n}{q^2}}\, 
  q^{(m+n)r}\, .
\end{array}
\end{equation}
To estimate \eq{App:special1}, we notice that
$$
\begin{array}{l}
\qbinom{x}{m}{q^2} \qbinom{y}{m}{q^2} \Big/
  \qbinom{x+r}{m+r}{q^2} \qbinom{y+r}{n+r}{q^2}
\vspace{2mm} \\
\qquad \displaystyle
  = \frac{\pn{x}}{\pn{x+r}} \cdot \frac{\pn{m+r}}{\pn{m}}
  \cdot \frac{\pn{y}}{\pn{y+r}} \cdot \frac{\pn{n+r}}{\pn{n}}
\end{array}
$$
This quantity is at most 1 (when $r=0$).
To get a lower bound we observe that the first and third ratios on the right
hand side are greater than 1, while the product of the second and third
is easily bounded
\begin{eqnarray*}
\frac{\pn{m+r}}{\pn{m}} \cdot \frac{\pn{n+r}}{\pn{n}} 
  &\geq& \prod_{k=1}^r (1 - q^{2(m+k)})^{-1} (1-q^{2(n+k)})^{-1} \\
  &\geq& \left(1 - \frac{q^{2(m+1)}}{1-q^2}\right)^{-1}
  \left(1 - \frac{q^{2(n+1)}}{1-q^2}\right)^{-1}\, .
\end{eqnarray*}
Inserting the inequality to \eq{App:special1}
$$
\begin{array}{l}
\displaystyle
\frac{ \ip{\psi^{+-}_{[1,x]}(m) \otimes \psi^{-+}_{[x+1,x+y+r]}(n+r)}
{\psi^{+-}_{[1,x+r]}(m+r) \otimes \psi^{-+}_{[x+r+1,x+y+r]}(n)} }
{\|\psi^{+-}_{[1,x]}(m) \otimes \psi^{-+}_{[x+1,x+y+r]}(n+r)\| \cdot
\|\psi^{+-}_{[1,x+r]}(m+r) \otimes \psi^{-+}_{[x+r+1,x+y+r]}(n)\|}
\vspace{2mm}  \\
\hspace{1cm} \displaystyle
\geq \left(1 - \frac{q^{2(m+1)}}{1-q^2}\right)^{-1/2}
  \left(1 - \frac{q^{2(n+1)}}{1-q^2}\right)^{-1/2}
\end{array}
$$
This leads to a useful formula. 
If $\psi$ and $\phi$ are normalized states then 
$\|\Proj(\psi) - \Proj(\phi)\| = \sqrt{1 - |\ip{\psi}{\phi}|^2}$.
Thus,
$$
\begin{array}{l}
\|\Proj(\psi^{+-}_{[1,x]}(m) \otimes \psi^{-+}_{[x+1,x+y+r]}(n+r))
  \vspace{2mm} \\
\displaystyle \hspace{25pt}
  - \Proj(\psi^{+-}_{[1,x+r]}(m+r) \otimes \psi^{-+}_{[x+r+1,x+y+r]}(n))\| 
  \leq \sqrt{\frac{8 q^2}{1 - q^2} (q^{2m} + q^{2n})}\, .
\end{array}
$$
In particular, changing notation to match the body of the paper,
\begin{equation}
\label{App:result1}
\|\Proj(\psi^{+-}_{[1,x]}(n_1)\otimes \psi^{-+}_{[x+1,L]}(n_2)) - 
  \Proj(\xi_{L,n_1+n_2}(\tilde{x}))\|
  \leq \frac{4 q^{\min(n_1,n_2)+1}}{\sqrt{1-q^2}}
\end{equation}
where $\tilde{x} = x + \floor{(n_2-n_1)/2}$.

To estimate \eq{App:special2}, we begin again by observing
\begin{eqnarray*}
&&\qbinom{x}{a}{q^2} \qbinom{y}{b}{q^2}
  \Big/ \qbinom{x+r}{m}{q^2} \qbinom{y+r}{n}{q^2}  \\
&& \hspace{75 pt}
  = \frac{\pn {x}}{\pn {x+r}} \cdot
  \frac{\pn {x-m+r}}{\pn {x-m}} \cdot
  \frac{\pn {y}}{\pn {y+r}} \cdot
  \frac{\pn{y-n+r}}{\pn{y-n}}\, .
\end{eqnarray*}
By the monotonicity of $\pn{x}$ in $x$, we have
$$
\pn \infty^2 \leq 
\qbinom{x}{m}{q^2} \qbinom{y}{n}{q^2}
  \Big/ \qbinom{x+r}{a}{q^2} \qbinom{y+r}{b}{q^2}  
\leq \frac{1}{f_q(\infty)^2}\, .
$$
From this it follows
\begin{equation}
\label{cf:n=0:bounds}
\begin{array}{l}
\displaystyle
\frac{\ip{\psi^{+-}_{[1,x]}(m) \otimes \psi^{-+}_{[x+1,x+y+r]}(n)}
{\psi^{+-}_{[1,x+r]}(m) \otimes \psi^{-+}_{[x+r+1,x+y+r]}(n)}}
{ \|\psi^{+-}_{[1,x]}(m) \otimes \psi^{-+}_{[x+1,x+y+r]}(n)\| 
\cdot
\|\psi^{+-}_{[1,x+r]}(m) \otimes \psi^{-+}_{[x+r+1,x+y+r]}(n)\| }
\vspace{2mm}  \\
\hspace{1cm} 
= C(x,y,m,n,r) q^{(m+n)r}\, ,
\end{array}
\end{equation}
where 
$$
f_q(\infty) \leq C(x,y,m,n,r) \leq \frac{1}{f_q(\infty)}\, .
$$
In particular, we have the useful bound
\begin{equation}
\label{App:result2}
\frac{|\ip{\xi_{L,n}(x)}{\xi_{L,n}(y)}|}
  {\|\xi_{L,n}(x)\| \cdot \|\xi_{L,n}(y)\|}
  \leq \frac{q^{n|y-x|}}{f_q(\infty)}\, .
\end{equation}
This is the first in a series of three inequalities needed for 
Section \ref{Sect:Eval}.

Next, we need a bound for 
$$
\frac{|\ip{\xi_{L,n}(x)}{H^{++}_{[1,L]} \xi_{L,n}(y)}|}
  {\|\xi_{L,n}(x)\| \cdot \|\xi_{L,n}(y)\|}.
$$
It turns out that the is well approximated by the normalized
inner-product above. 
The reason is that, while $H^{++}_{[1,L]}$ is not a 
small operator in general, when acting on the droplet states
it reduces to just one nearest-neighbor interaction:
$H^{++}_{[1,L]} \xi_{L,n}(x) = H^{++}_{x,x+1} \xi_{L,n}(x)$.
To exploit this we return to the notation above, and observe 
that as long as $r\geq 1$
\begin{equation}
\label{App:prelim3}
\begin{array}{l}
\displaystyle
\ip{\psi^{+-}_{[1,x]}(m) \otimes \psi^{-+}_{[x+1,x+y+r]}(n+k)}
{\psi^{+-}_{[1,x+r]}(m+k) \otimes \psi^{-+}_{[x+r+1,x+y+r]}(n)}
\vspace{2mm}  \\
\hspace{1cm} \displaystyle
= \sum_{j,l} q^{(r+2) m + n + k - 3 j + (r - 2) l}
  \ip{\psi^{+-}_{[1,x-1]}(m-j)}{\psi^{+-}_{[1,x-1]}(m-j)} \vspace{2mm}\\
\hspace{1cm} \displaystyle
  \times \ip{\psi^{+-}_{\{x\}}(j) \otimes \psi^{-+}_{\{x+1\}}(l)}
  {\psi^{+-}_{[x,x+1]}(j+l)} \vspace{2mm}\\
\hspace{1cm} \displaystyle
  \times \ip{\psi^{-+}_{[x+12,x+y+r]}(n+k-l)}
  {\psi^{+-}_{[x+2,x+r]}(k-l) \otimes 
  \psi^{-+}_{[x+r+1,x+y+r]}(n)}\, .
\end{array}
\end{equation}
This is derived just as before, using equations \eq{App:+-coprod} --
\eq{App:mixed-ip}. 
Note 
$$
\psi^{+-}_{\{x\}}(j) = \psi^{-+}_{\{x\}}(j) = q^j (S_x^{-})^j 
\ket{\uparrow}_x\, .
$$
The usefulness of this formula is in the fact that 
\begin{eqnarray*}
&& |\ip{\psi^{+-}_{\{x\}}(j) \otimes \psi^{-+}_{\{x+1\}}(l)}
  {H^{++}_{x,x+1} \psi^{+-}_{[x,x+1]}(j+l)}| \\
&& \qquad \leq \ip{\psi^{+-}_{\{x\}}(j) \otimes \psi^{-+}_{\{x+1\}}(l)}
  {\psi^{+-}_{[x,x+1]}(j+l)}\, .
\end{eqnarray*}
Indeed, the formula for the right-hand-side
is
$$
\ip{\psi^{+-}_{\{x\}}(j) \otimes \psi^{-+}_{\{x+1\}}(l)}
  {\psi^{+-}_{[x,x+1]}(j+l)}
  = q^{2j+3l}\, ,
$$
while the left-hand-side is
$$
\begin{array}{c|c|l}
\displaystyle
j & l & \ip{\psi^{+-}_{\{x\}}(j) \otimes \psi^{-+}_{\{x+1\}}(l)}
  {H^{++}_{x,x+1} \psi^{+-}_{[x,x+1]}(j+l)}  \\
\hline
\displaystyle
\rule{0mm}{5mm} 
0 & 0 & - A(\Delta) \\
\rule{0mm}{6mm} 
\displaystyle
0 & 1 & \displaystyle  \frac{q^2 (1-q)^2}{2(1+q^2)} \\
\rule{0mm}{6mm} 
\displaystyle
1 & 0  & \displaystyle - \frac{q^4(1-q^2)}{2(1+q^2)} \\
\rule{0mm}{5mm} 
\displaystyle
1 & 1 & A(\Delta) q^5
\end{array}
$$
Thus,
$$
\begin{array}{l}
\displaystyle
|\ip{\psi^{+-}_{[1,x]}(m) \otimes \psi^{-+}_{[x+1,x+y+r]}(n+k)}
{H^{++}_{x,x+1} \psi^{+-}_{[1,x+r]}(m+k) \otimes \psi^{-+}_{[x+r+1,x+y+r]}(n)}|
\vspace{2mm}  \\
\hspace{0.5cm} \displaystyle
\leq \sum_{j,l} q^{(r+2) m + n + k - 3 j + (r - 2) l}
  \ip{\psi^{+-}_{[1,x-1]}(m-j)}{\psi^{+-}_{[1,x-1]}(m-j)} \vspace{2mm}\\
\hspace{1.0cm} \displaystyle
  \times |\ip{\psi^{+-}_{\{x\}}(j) \otimes \psi^{-+}_{\{x+1\}}(l)}
  {H^{++}_{x,x+1} \psi^{+-}_{[x,x+1]}(j+l)}| \vspace{2mm}\\
\hspace{1.0cm} \displaystyle
  \times \ip{\psi^{-+}_{[x+12,x+y+r]}(n+k-l)}
  {\psi^{+-}_{[x+2,x+r]}(k-l) \otimes 
  \psi^{-+}_{[x+r+1,x+y+r]}(n)} \vspace{2mm}\\
\hspace{0.5cm} \displaystyle
\leq \sum_{j,l} q^{(r+2) m + n + k - 3 j + (r - 2) l}
  \ip{\psi^{+-}_{[1,x-1]}(m-j)}{\psi^{+-}_{[1,x-1]}(m-j)} \vspace{2mm}\\
\hspace{1.0cm} \displaystyle
  \times {\psi^{+-}_{\{x\}}(j) \otimes \psi^{-+}_{\{x+1\}}(l)}
  {\psi^{+-}_{[x,x+1]}(j+l)} \vspace{2mm}\\
\hspace{1.0cm} \displaystyle
  \times \ip{\psi^{-+}_{[x+12,x+y+r]}(n+k-l)}
  {\psi^{+-}_{[x+2,x+r]}(k-l) \otimes 
  \psi^{-+}_{[x+r+1,x+y+r]}(n)} \vspace{2mm}\\
\hspace{0.5cm} \displaystyle 
  = \ip{\psi^{+-}_{[1,x]}(m) \otimes \psi^{-+}_{[x+1,x+y+r]}(n+k)}
  {\psi^{+-}_{[1,x+r]}(m+k) \otimes \psi^{-+}_{[x+r+1,x+y+r]}(n)}\, .
\end{array}
$$
This result, in conjunction with \eq{App:result2}, gives
\begin{equation}
\label{App:result3}
\frac{|\ip{\xi_{L,n}(x)}{H^{++}_{[1,L]} \xi_{L,n}(y)}|}
  {\|\xi_L(x,n)\| \cdot \|\xi_L(y,n)\|}
  \leq \frac{q^{n|y-x|}}{f_q(\infty)}\, ,
\end{equation}
whenever $|x-y|\geq 1$.
The requirement that $|x-y|\geq 1$ comes from the fact that $r$ must be 
at least one for \eq{App:prelim3} to hold true.

Now a similar argument works to bound 
$|\ip{\xi_{L,n}(x)}{(H^{++}_{[1,L]})^2 \xi_{L,n}(y)}|$
by $\ip{\xi_{L,n}(x)}{\xi_{L,n}(y)}$.
Specifically, we note
$$
\ip{\xi_{L,n}(x)}{(H^{++}_{[1,L]})^2 \xi_{L,n}(y)}
  = \ip{\xi_{L,n}(x)}{(H^{++}_{x,x+1} H^{++}_{y,y+1} \xi_{L,n}(y)}
$$
as long as $|x-y| \geq 2$.
Then the same argument as above can show that
$$
|\ip{\xi_{L,n}(x)}{(H^{++}_{x,x+1} H^{++}_{y,y+1} \xi_{L,n}(y)}|
  \leq \ip{\xi_{L,n}(x)}{\xi_{L,n}(y)}\, .
$$
Thus we have
\begin{equation}
\label{App:result4}
\frac{|\ip{\xi_{L,n}(x)}{(H^{++}_{[1,L]})^2 \xi_{L,y}(n)}|}
  {\|\xi_{L,n}(x)\| \cdot \|\xi_{L,y}(n)\|}
  \leq \frac{q^{n|y-x|}}{f_q(\infty)}\, ,
\end{equation}
whenever $|x-y|\geq 2$.

%%%%%%%%%%%%%%%%%%%%%%%%%%%%%%%%%%%%%%%%%%%%%%%%%%%%%%%%%%%%%%%%%%%
%%%%%%%%%%%%%%%%%%%%%%%%%%%%%%%%%%%%%%%%%%%%%%%%%%%%%%%%%%%%%%%%%%%
%%%%%%%%%%%%%%%%%%%%%%%%%%%%%%%%%%%%%%%%%%%%%%%%%%%%%%%%%%%%%%%%%%%

\section*{Appendix B}
\label{Sec:AppB}
\setcounter{equation}{0}
\renewcommand{\theequation}{B.\arabic{equation}}

In this section we derive a single result.
We need the following definitions, some of which appeared previously
in the paper.
Given an arbitrary finite subset $\Lambda \subset \Ir$,
let $\Hil_\Lambda$ be the $|\Lambda|$-fold
tensor product $\bigotimes_{x\in \Lambda} \Cx_x^2$,
the space of all spin states on $\Lambda$.
The subspace of all vectors $\psi \in \Hil_\Lambda$ with exactly $n$
down spins is denoted $\Hil_{\Lambda,n}$.
For any subset $\Lambda_1 \subset \Lambda$,
we can define $Q_{\Lambda_1,n}$ to be the projection onto the subspace of 
$\Hil_\Lambda$ consisting of those vectors with exactly $n$ down spins in 
$\Lambda_1$.
So, $Q_{\Lambda_1,n} = \Proj(\Hil_{\Lambda_1,n} \otimes 
\Hil_{\Lambda\setminus \Lambda_1})$.
We also define $P_{\Lambda_1} = Q_{\Lambda_1,0} + Q_{\Lambda_1,|\Lambda_1|}$. 
It is the projection onto the span of vectors such that on $\Lambda_1$ they 
have all up spins or all down spins, but nothing else.

Now, let $0 \leq n < L$. 
Suppose $J = [a,b]$ is a subinterval of $[1,L]$.
We define the projections:
\begin{eqnarray*}
  G^\uparrow_j &=& Q_{[1,a-1],j}\, Q_{J,0}\, Q_{[b+1,L],n-j}\, ,\\
  G^\downarrow_j &=& Q_{[1,a-1],j}\, Q_{J,|J|}\, Q_{[b+1,L],n-j-|J|}\, .
\end{eqnarray*}
Then, for any $\psi \in \Hil_{[1,L],n}$,
$$
P_J \psi = \sum_{j=0}^n G^\uparrow_j \psi 
  + \sum_{j=0}^{n-|J|} G^\downarrow_j \psi\, .
$$
We recall the definition of droplet states:
For $\floor{n/2} \leq x \leq L-\ceil{n/2}$, 
$$
\xi_{L,n}(x) 
  = \psi^{+-}_{[1,x]}(\floor{n/2}) \otimes \psi^{-+}_{[x+1,L]}(\ceil{n/2})\, ,
$$
where $\psi^{+-}_{[1,x]}(\floor{n/2})$ and $\psi^{-+}_{[x+1,L]}(\ceil{n/2})$
are the kink and antikink states defined in \eq{Intro:+-}
and \eq{Intro:-+}.
Let $\Xi_{x} = \Proj(\xi_{L,n}(x))$.
Define the intervals 
\begin{eqnarray*}
I_1 &=& [\floor{n/2},a-\ceil{n/2}-1]\, ,\\
I_2 &=& [b-\ceil{n/2},a-1+\floor{n/2}]\, ,\\  
I_3 &=& [b+\floor{n/2},L-\ceil{n/2}]\, .
\end{eqnarray*}
Some of these intervals may be empty.
We have the following result.
There exists an $N(q) \in \Nl$ and a $C(q) < \infty$,
such that as long as $n \geq N(q)$
\begin{eqnarray*}
\sum_{x \in I_1 \cup I_2 \cup I_3} P_J \Xi_x P_J
  &\geq&  \sum_{x \in I_1} G^\uparrow_n \Xi_x G^\uparrow_n
  + \sum_{x \in I_2} G^\downarrow_{a-1+\floor{n/2}-x} \Xi_x 
  G^\downarrow_{a-1+\floor{n/2}-x} \\
  &&+ \sum_{x \in I_3} G^\uparrow_0 \Xi_x G^\uparrow_0
  -C(q) q^{|J|} P_J \Proj(\Hil_{[1,L],n}) \, .
\end{eqnarray*}

To prove this we group certain projections, $G_j^\sigma$,
and certain projections, $\Xi_x$, together.
Let
$$
\begin{array}{rclrcl}
\displaystyle
\mathcal{G}_1 &=
&\displaystyle \sum_{j=0}^{n-|J|} G^\downarrow_j\, ,
&\displaystyle \mathcal{X}_1 &=
&\displaystyle \sum_{j=0}^{n-|J|} \Xi_{a-1+\floor{n/2}-j}\, ;
\vspace{1mm}\\
\displaystyle \mathcal{G}_2 &= 
&\displaystyle G^\uparrow_0\, ,
&\displaystyle \mathcal{X}_2 &= 
&\displaystyle \sum_{x=b+\floor{n/2}}^{L-\ceil{n/2}} \Xi_x\, ;
\vspace{1mm}\\
\mathcal{G}_3 &=
&\displaystyle  \sum_{j=1}^{\floor{n/2}-1} G^\uparrow_j\, ;
\vspace{1mm}\\
\mathcal{G}_4 &=
&\displaystyle  G^\uparrow_{\floor{n/2}}\, ;
\vspace{1mm}\\
\mathcal{G}_5 &= 
&\displaystyle \sum_{j=\floor{n/2}+1}^{n-1} G^\uparrow_j\, ;
\vspace{1mm}\\
\mathcal{G}_6 &= 
&\displaystyle G^\uparrow_n\, ,
&\displaystyle \mathcal{X}_6 &= 
&\displaystyle \sum_{x=\floor{n/2}}^{a-1-\ceil{n/2}} \Xi_x\, .
\end{array}
$$
To prove the claim it suffices to prove 
$\|\mathcal{X}_i \mathcal{G}_j\| \leq O(q^{|J|})$ for 
$i \neq j$, and 
\begin{equation}
\label{AppB:1}
\| \mathcal{G}_1 \mathcal{X}_1 \mathcal{G}_1 
  - \sum_{j=0}^{n-|J|} G^\downarrow_j \cdot \Xi_{a-1+\floor{n/2}-j} 
  \cdot G^\downarrow_j \|
  \leq O(q^{|J|})\, .
\end{equation}
We will explain how this may be done now.

By our definition, each $\mathcal{G}_i$ may be written
$\sum_{k \in E_i} G^{\sigma_i}_k$, and each $\mathcal{X}_j$ 
may be written $\sum_{x \in F_j} \Xi_x$, for intervals
$E_i, F_j$, possibly empty, and $\sigma_i \in \{\uparrow,\downarrow\}$.
Thus, letting $\sigma = \sigma_i$,
\begin{eqnarray*}
&& (\mathcal{X}_j \mathcal{G}_i)^* (\mathcal{X}_j \mathcal{G}_i)
  \, =\,  \sum_{k,l \in E_i} \sum_{x,y \in F_j}
  G^\sigma_k \Xi_x \Xi_y G^\sigma_l \\
&& \quad =  \sum_{k,l \in E_i} \sum_{x,y \in F_j}
  G^\sigma_k \cdot
  \frac{\ket{\xi_{L,n}(x)}\bra{\xi_{L,n}(x)}}
  {\ip{\xi_{L,n}(x)}{\xi_{L,n}(x)}} \cdot
  \frac{\ket{\xi_{L,n}(y)}\bra{\xi_{L,n}(y)}}
  {\ip{\xi_{L,n}(y)}{\xi_{L,n}(y)}} \cdot
  G^\sigma_l \\
&& \quad =  \sum_{k,l \in E_i} \sum_{x,y \in F_j}
  G^\sigma_k \cdot
  \frac{\ket{G^\sigma_k \xi_{L,n}(x)}}{\|\xi_{L,n}(x)\|} \cdot
  \frac{\ip{\xi_{L,n}(x)}{\xi_{L,n}(y)}}
  {\|\xi_{L,n}(x)\| \cdot \|\xi_{L,n}(y)\|} \cdot
  \frac{\bra{G^\sigma_l \xi_{L,n}(y)}}{\|\xi_{L,n}(y)\|}
  \cdot G^\sigma_l
\end{eqnarray*}
Applying Cauchy-Schwarz we deduce that 
$$
\| \mathcal{X}_j \mathcal{G}_i \psi \|^2
  \leq \sum_{k,l \in E_i} \|G_k \psi\| \, \|G_l \psi\|  
  M^{j \sigma_i}_{kl}\, ,
$$
where 
$$
M^{j \sigma}_{kl} = \sum_{xy \in F_j}
  \frac{\|G^\sigma_k \xi_{L,n}(x)\|}{\|\xi_{L,n}(x)\|} \cdot
  \frac{|\ip{\xi_{L,n}(x)}{\xi_{L,n}(y)}|}
  {\|\xi_{L,n}(x)\| \cdot \|\xi_{L,n}(y)\|} \cdot
  \frac{\|G^\sigma_l \xi_{L,n}(y)\|}{\|\xi_{L,n}(y)\|}\, .
$$
Since the projections $G^\sigma_k$ are mutually orthogonal to
one another, $\|\mathcal{G}_i \psi\|^2 
= \sum_{k \in E_i} \|G_k^\sigma \psi\|^2$.
Thus,
$$
\| \mathcal{X}_j \mathcal{G}_i \psi \|^2
  \leq \|\mathcal{G}_i \psi\|^2 \cdot 
  \|(M^{j \sigma_i}_{kl \in E_i})_{kl}\|\, .
$$
Of course, $\|\mathcal{G}_i \psi\|^2 \leq \|\psi\|^2$,
because $\mathcal{G}_i$ is a projection.
So 
$$
\| \mathcal{X}_j \mathcal{G}_i \| \leq
\|(M^{j \sigma_i}_{kl})_{kl \in E_i}\|^{1/2}\, .
$$

We now discuss how to bound $\|(M^{j \sigma_i}_{kl})_{kl \in E_i}\|$.
We can bound the inner-product 
$\ip{\xi_{l,n}(x)}{\xi_{l,n}(y)}$ by \eq{App:result1}.
So
$$
M^{j \sigma}_{kl} 
  \leq \sum_{xy \in F_j} \frac{q^{n|x-y|}}{\pn{\infty}}
  \cdot \frac{\|G^\sigma_k \xi_{L,n}(x)\|}{\|\xi_{L,n}(x)\|} \cdot
  \frac{\|G^\sigma_l \xi_{L,n}(y)|}{\|\xi_{L,n}(y)\|}\, .
$$
Then, using the operator norm with respect $l^\infty$,
\begin{eqnarray*}
\|(M^{j \sigma}_{kl})_{kl \in E_i}\|
  &\leq& \|(M^{j \sigma}_{kl})_{kl \in E_i}\|_\infty \\
  &\leq& 
  \sup_{k \in E_i}\, \sum_{l \in E_i} \sum_{x,y \in F_j}
  \frac{q^{n|x-y|}}{\pn{\infty}}
  \cdot \frac{\|G^\sigma_k \xi_{L,n}(x)\|}{\|\xi_{L,n}(x)\|} \cdot
  \frac{\|G^\sigma_l \xi_{L,n}(y)|}{\|\xi_{L,n}(y)\|}\, .
\end{eqnarray*}
To proceed, we need to estimate 
$\|G^\sigma_l \xi_{L,n}(x)\|/\|\xi_{L,n}(x)\|$
for each $\sigma$, $l$ and $x$.
In fact, no estimation is required, we can perform the computation 
exactly.
Let us explain how this is done.
The operator $G^\sigma_l$ falls in the following class of projections.
Suppose we have some partition $\mathcal{P}$ of $[1,L]$,
composed of intervals $[x_{j-1}+1,x_j]$ where $0=x_0<x_1<\dots<x_r=L$,
and suppose we have a vector $\vec{n} = (n_1,\dots,n_r)$,
where $0 \leq n_j \leq x_j-x_{j-1}$ and $\sum_{j=1}^r n_j = n$.
Then we can define the projection 
$$
Q_{\mathcal{P},\vec{n}} := \prod_{j=1}^r Q_{[x_{j-1}+1,x_j],n_j}\, .
$$  
The operators $G^\sigma_l$ are of this form, where the partition
has three intervals $[1,a-1]$, $[a,b]$ and $[b+1,L]$, and 
$\vec{n} = (j,0,n-j)$ or $\vec{n}=(j,|J|,n-j-|J|)$, depending on whether
$\sigma$ is $\uparrow$ or $\downarrow$.
We can reduce the problem of computing 
$Q_{\mathcal{P},\vec{n}} \xi_{L,n}(x)$ to one of computing
$Q_{\mathcal{P}_1,\vec{n}_1} \psi^{+-}_{[1,x]}(\floor{n/2})$,
and
$Q_{\mathcal{P}_2,\vec{n}_2} \psi^{+-}_{[x+1,L]}(\ceil{n/2})$
for some partitions and vectors $\mathcal{P}_1$,$\mathcal{P}_2$,
$\vec{n}_1$ and $\vec{n}_2$.
To accomplish this, let $k$ be the integer such that $x_{k-1}+1 \leq x < x_k$.
Define the partition $\mathcal{P}'$ where $x_j'=x_j$ for $j<k$,
$x_k = x$, and $x'_j = x_{j-1}$ for $j>k$, and
define the $r+1$-vector $\vec{n}'$ by
$n'_j = n_j$ for $j <k$, $n'_k = \floor{n/2} - \sum_{j=1}^{k-1} n_j$,
$n_{k+1} = n_k - n'_k$, and $n'_j = n_{j-1}$ for $j > k+1$.
Since $\xi_{L,n}(x)$ has a definite number of downspins,
$\floor{n/2}$, to the left of $x$ and a definite number of 
downspins, $\ceil{n/2}$, to the right of $x+1$, the vector 
$Q_{\mathcal{P},\vec{n}} \xi_{L,n}(x)$ is the same as
$Q_{\mathcal{P}',\vec{n}'} \xi_{L,n}(x)$.
In fact, since $\xi_{L,n}(x) = \psi^{+-}_{[1,x]}(\floor{n/2}) \otimes 
\psi^{-+}_{[x+1,L]}(\ceil{n/2})$, we know
$$
Q_{\mathcal{P},\vec{n}} \xi_{L,n}(x) 
  = (Q_{\mathcal{P}_1,\vec{n}_1} \psi^{+-}_{[1,x]}(\floor{n/2}))
  \otimes (Q_{\mathcal{P}_2,\vec{n}_2} \psi^{-+}_{[x+1,L]}(\ceil{n/2}))\, ,
$$
where $\mathcal{P}_1$ is the partition consisting of the first $k$
parts of $\mathcal{P}'$,
$\mathcal{P}_2$ is the remainder partition, 
$\vec{n}_1 = (n'_1,\dots,n'_k)$ and 
$\vec{n}_2 = (n'_{k+1},\dots,n'_r)$.
Therefore,
$$
\frac{\|Q_{\mathcal{P},\vec{n}} \xi_{L,n}(x)\|}
{\|\xi_{L,n}(x)\|}
  = \frac{\|Q_{\mathcal{P}_1,\vec{n}_1} \psi^{+-}_{[1,x]}(\floor{n/2})\|}
  {\|\psi^{+-}_{[1,x]}(\floor{n/2})\|} \cdot
  \frac{Q_{\mathcal{P}_2,\vec{n}_2} \psi^{-+}_{[x+1,L]}(\ceil{n/2})\|}
  {\|\psi^{-+}_{[x+1,L]}(\ceil{n/2})\|}\, .
$$
We now present the formula for the two quantities on the right-hand-side 
of the equation.

The key to the computation is the decomposition formulae of
\eq{App:+-coprod} and \eq{App:-+coprod}.
These have trivial generalizations.
Specifically, for $x_0 < x_1 < \dots <x_r$, 
\begin{eqnarray}
\label{CB+-:gen}
\psi^{+-}_{[x_0+1,x_r]}(n)
  &=& \sum_{(n_1,\dots,n_r) \atop n_1+\dots+n_r=n}
  q^{n x_r - (n_1 x_1 + \dots n_r x_r)}
  \bigotimes_{j=1}^r \psi^{+-}_{[x_{j-1}+1,x_j]}(n_j) \\
\label{CB-+:gen}
\psi^{-+}_{[x_0+1,x_r]}(n)
  &=& \sum_{(n_1,\dots,n_r) \atop n_1+\dots+n_r=n}
  q^{(n_1 x_0 + \dots n_r x_{r-1}) - n x_0}
  \bigotimes_{j=1}^r \psi^{-+}_{[x_{j-1}+1,x_j]}(n_j) 
\end{eqnarray}
From this one can easily calculate
\begin{eqnarray}
\label{projform+-}
\frac{\|Q_{\mathcal{P},\vec{n}} \psi^{+-}_{[x_0+1,x_r]}(n)\|^2}
{\|\psi^{+-}_{[x_0+1,x_r]}(n)\|^2}
  = \frac{\prod_{j=1}^r \qbinom{x_j - x_{j-1}}{n_j}{q^2}}
  {\qbinom{x_r-x_0}{n}{q^2}}\, 
  q^{\sum_{j=1}^r n_j(2(x_r-x_j)-(n-n_j))} \\
\label{projform-+}
\frac{\|Q_{\mathcal{P},\vec{n}} \psi^{-+}_{[x_0+1,x_r]}(n)\|^2}
{\|\psi^{-+}_{[x_0+1,x_r]}(n)\|^2}
  = \frac{\prod_{j=1}^r \qbinom{x_j - x_{j-1}}{n_j}{q^2}}
  {\qbinom{x_r-x_0}{n}{q^2}}\, 
  q^{\sum_{j=1}^r n_j(2(x_{j-1}-x_0)-(n-n_j))} \\
\end{eqnarray}
We notice the following interesting fact.
The exponent of $q$ in the formulas above has the following 
interpretation.
The most probable locations of the downspins for 
kink state $\psi^{+-}_{[1,L]}(n)$ are in the interval $[L+1-n,L]$.
Suppose we place marbles in these places and ask for
the minimum transport required to move these marbles
so that $n_j$ of the marbles lie in the 
bin $[x_{j-1}+1,x_j]$ for each $j$.
Then this is precisely the exponent of $q$ in \eq{projform+-}.
To state this in symbols
\begin{eqnarray*}
&&\sum_{j=1}^r n_j(2(x_r - x_1) - (n-n_j)) = 
\min\{\sum_{x=1}^L |f(x) - x| : f \in \textrm{Perm}([1,L]), \\ 
&&\qquad \#\big(f([L+1-n,L]) \cap [x_{j-1}+1,x_j]\big) = n_j, j=1,\dots,r\} 
\end{eqnarray*}
The exponent of $q$ in \eq{projform-+} has a similar interpretation,
except that the marbles initially occupy the sites of
$[1,n]$ instead of $[L+1-n,L]$.

Having said how one can perform the computations of 
$\|G^\sigma_j \xi_{L,n}(x)\|$, we now state our results.
The following notation is convenient:
$$
\EXP{*}{L,n,x} := \frac{\ip{\xi_{L,n}(x)}{* \xi_{L,n}(x)}}
  {\ip{\xi_{L,n}(x)}{\xi_{L,n}(x)}}\, .
$$
This is the expectation value of an observable with respect to
the droplet state $\xi_{L,n}(x)$. 
\begin{itemize}
\item
If $0 \leq x \leq a-1$ and $\sigma = \uparrow$ let 
$r = a-1-x-j+\floor{n/2}$. Then
$$
\EXP{G^\uparrow_j}{L,n,x}
  = \frac{\qbinom{a-1-x}{r}{q^2} \qbinom{L-b}{n-j}{q^2}}
  {\qbinom{L-x}{\ceil{n/2}}{q^2}} q^{2(n-j)(|J|+r)}\, .
$$
We make the convention that 
$$
\qbinom{n}{k}{q^2} = 0
\quad
\textrm{if}
\quad k<0 
\quad
\textrm{or}
\quad k>n\, .
$$
Thus the formula above is zero unless $0\leq r\leq a-1-x$.
\item
If $0 \leq x \leq a-1$ and $\sigma = \downarrow$ let 
$r = a-1-x-j+\floor{n/2}$. Then
$$
\EXP{G^\downarrow_j}{L,n,x}
  = \frac{\qbinom{a-1-x}{r}{q^2} \qbinom{L-b}{n-j-|J|}{q^2}}
  {\qbinom{L-x}{\ceil{n/2}}{q^2}} q^{2(n-j)r}\, .
$$
\item
If $a \leq x \leq b$ and $\sigma = \uparrow$, 
the answer is zero unless $j = \floor{n/2}$, and
$$
\EXP{G^\uparrow_{\floor{n/2}}}{L,n,x}
  = \frac{\qbinom{a-1}{\floor{n/2}}{q^2} \qbinom{L-b}{\ceil{n/2}}{q^2}}
  { \qbinom{x}{\floor{n/2}}{q^2} \qbinom{L-x}{\ceil{n/2}}{q^2} }
  q^{2[\floor{n/2}(x-a+1) + \ceil{n/2}(b-x)]}\, .
$$
\item
If $a \leq x \leq b$ and $\sigma = \downarrow$, the answer is zero
unless $j = \floor{n/2}-x+a-1$, and
$$
\EXP{G^\downarrow_{\floor{n/2}}}{L,n,x}
  = \frac{\qbinom{a-1}{x-\floor{n/2}}{q^2} \qbinom{L-b}{L-x-\ceil{n/2}}{q^2}}
  { \qbinom{x}{\floor{n/2}}{q^2} \qbinom{L-x}{\ceil{n/2}}{q^2} }\, .
$$
\item
If $b+1 \leq x \leq L$ and $\sigma = \uparrow$, let 
$r = x-b-\floor{n/2}+j$.
Then
$$
\EXP{G^\uparrow_j}{L,n,x}
  = \frac{\qbinom{a-1}{j}{q^2} \qbinom{x-b}{r}{q^2}}
  {\qbinom{x}{\floor{n/2}}{q^2}} q^{2j(|J|+r)}\, .
$$
\item
If $b+1\leq x\leq L$ and $\sigma = \downarrow$, let
$x-a+1-\floor{n/2}+j$.
Then
$$
\EXP{G^\uparrow_j}{L,n,x}
  = \frac{\qbinom{a-1}{j}{q^2} \qbinom{x-b}{r}{q^2}}
  {\qbinom{x}{\floor{n/2}}{q^2}} q^{2j(|J|+r)}\, .
$$
\end{itemize}

The rest of the computations proceed directly from these observations.
Note that each $q^2$-binomial coefficient can be bounded above
by $\pn \infty ^{-1}$, but one should remember to restrict the
indices $j$ and $x$ to those for which none of the 
$q^2$-binomial coefficients vanish.
Our results are the following:
\begin{itemize}
\item
As mentioned above, it is easy to check that 
$$
\mathcal{X}_1 \mathcal{G}_2 
  = \mathcal{X}_1 \mathcal{G}_6 
  = \mathcal{X}_2 \mathcal{G}_6 
  = \mathcal{X}_2 \mathcal{G}_5 
  = \mathcal{X}_6 \mathcal{G}_2 
  = \mathcal{X}_6 \mathcal{G}_3
  = 0\ .
$$
Simply put, if one consults the formulae in the paragraph,
each of the products above is composed of $\Xi_x G^\sigma_j$
for which the $q$-binomial coefficients vanish.
\item
A simultaneous bound for $\|\mathcal{X}_1 \mathcal{G}_3\|^2$
and $\|\mathcal{X}_1 \mathcal{G}_5\|^2$ is $C(q) q^{2|J|}$,
where
$$
C(q) = \frac{2+8q}{(1-q)^4 \pn{\infty}^3}\, .
$$
\item 
$$\|\mathcal{X}_1 \mathcal{G}_4\|^2
  \leq \frac{1}{\pn{\infty}^3}
  \left(|J| + \frac{1+q^{\floor{n/2}}}{1-q^{\floor{n/2}}}\right)^2
  q^{2 |J| \floor{n/2}}\, .
$$
\item
We bound $\|\mathcal{X}_2 \mathcal{G}_1\|^2$
and $\|\mathcal{X}_6 \mathcal{G}_1\|^2$, simultaneously,
by $C(q) q^{2(|J|-1)^2}$, where
$$
C(q) = \frac{1}{\pn{\infty}^3(1-q^{|J|})^2 (1-q^{2(|J|-1)})}\, .
$$
The reason the bound is so small is that it is actually equal to zero,
if $|J| > n$,
as can be understood by counting downspins to the left and right of
$x$.
\item
Both 
$\|\mathcal{X}_2 \mathcal{G}_3\|^2$ and
$\|\mathcal{X}_6 \mathcal{G}_5\|^2$ can each be bounded
by $C(q) q^{2|J|}$, where
$$
C(q) = \frac{q^2}{\pn{\infty}^3 (1 - q)^2 (1 - q^{|J|+2})}\, .
$$
\item
Both $\|\mathcal{X}_2 \mathcal{G}_4\|^2$ and
$\|\mathcal{X}_6 \mathcal{G}_4\|^2$ can each be bounded
by 
$$
\frac{1}{\pn{\infty}^3} \left( \frac{1+q^{2\ceil{n/2}}}
  {1 - q^{2\ceil{n/2}}} + \frac{1+q^{2\floor{n/2}}}{1-q^{2\floor{n/2}}}
  \right)
q^{4 \ceil{n/2} (|J| + \ceil{n/2})}\, .
$$
\end{itemize}
That accounts for all of the necessary computations except one,
which we now carry out.

We show in this paragraph that
\begin{equation}
\label{example}
\left\|\mathcal{G}_1\mathcal{X}_1 \mathcal{G}_1 - 
  \sum_{j=0}^{n-|J|} G^\downarrow_{j} 
  \Xi_{a-1+\floor{n/2}+j} G^\downarrow_{j}
  \right\| \leq \frac{4 q^{|J|}}{\pn{\infty}^3 (1 - q^{|J|})^2}\, .
\end{equation}
In this case we can define $x_j = a-1+\floor{n/2}+j$, 
for each $0 \leq j \leq n-|J|$, and we have
$$
\EXP{G_j^\downarrow}{L,n,x} 
  \leq \frac{1}{\pn{\infty}} q^{|J|\cdot|x-x_j|}\, .
$$
This is understood because $|x-x_j|$ downspins must be moved all the way 
across the droplet in order to change the basic interval for
$\xi(x)$ into a state compatible with $\mathcal{G}^\downarrow_j$.
Thus, proceeding in the same way as before, we obtian
$$
\left\|\mathcal{G}_1\mathcal{X}_1 \mathcal{G}_1 - 
  \sum_{j=0}^{n-|J|} G^\downarrow_{j} 
  \Xi_{a-1+\floor{n/2}+j} G^\downarrow_{j}
  \right\| \leq \|\mathcal{M}\|\ .
$$
where $\mathcal{M}_{jj}=0$ for each $j$, and
$$
\mathcal{M}_{jk} \leq \frac{1}{\pn{\infty}^2} \sum_{x\in I_2}
  q^{|J|\cdot |x-x_j| + |J|\cdot |x-x_k|}
$$
when $j \neq k$.
By extending the indices $x$ to cover all integers, and
by translating so that $x_j$ is the new origin of $x$,
we have
$$
\mathcal{M}_{jk} \leq \frac{1}{\pn{\infty}^3} \sum_{x}
  q^{|J|\cdot |x| + |J|\cdot |x+j-k|}\, .
$$
The series is easily calculated as
$$
\sum_{x}
  q^{|J|\cdot |x| + |J|\cdot |x+j-k|}
  = q^{|J|\cdot|j-k|} \left(|j-k| + \frac{1+q^{2|J}}{1-q^{2|J|}}\right)\, .
$$
So, for any fixed $j$, we have
$$
\sum_{k\in \Ir \atop k \neq j} \mathcal{M}_{jk}
  \leq \frac{2}{\pn{\infty}^2}
  \sum_{l=1}^\infty 
  q^{|J|l} \left(l + \frac{1+q^{2|J}}{1-q^{2|J|}}\right)\, .
$$
This sum is then easily computed as
$$
\sum_{l=1}^\infty 
  q^{|J|l} \left(l + \frac{1+q^{2|J|}}{1-q^{2|J|}}\right)
  = \frac{4 q^{|J|}}{(1-q^{|J|})^2}\, .
$$
From this we obtain \eq{example}.  

%%%%%%%%%%%%%%%%%%%%%%%%%%%%%%%%%%%%%%%%%%%%%%%%%%%%%%%%%%%%%%%%%%%
%%%%%%%%%%%%%%%%%%%%%%%%%%%%%%%%%%%%%%%%%%%%%%%%%%%%%%%%%%%%%%%%%%%
%%%%%%%%%%%%%%%%%%%%%%%%%%%%%%%%%%%%%%%%%%%%%%%%%%%%%%%%%%%%%%%%%%%

\section*{Acknowledgements}

This material is based on work supported by the National Science Foundation 
under Grant No. DMS0070774.

\end{document}